\documentclass[aps,prc,twocolumn,superscriptaddress,showpacs,showkeys,amsmath,amssymb,nofootinbib]{revtex4-2}

\usepackage[T1]{fontenc}
\usepackage[utf8]{inputenc}

\usepackage{amsmath,amssymb}
\usepackage{bm}
\usepackage{physics}
\usepackage{graphicx}
\usepackage{float}
\usepackage{booktabs}
\usepackage{enumitem}
\usepackage{xcolor}
\usepackage{array}
\usepackage[normalem]{ulem}

\usepackage[unicode=true,colorlinks=true,
  citecolor=blue,linkcolor=blue,urlcolor=blue]{hyperref}

\begin{document}

\title{Four-neutron halo model at the unitary limit}

\author{R.~M.~Francisco}
\affiliation{Instituto Tecnol\'{o}gico de Aeron\'{a}utica,
Pra\c{c}a Mal.\ Eduardo Gomes, 12228-900, S\~{a}o Jos\'{e} dos Campos, SP, Brazil}
\affiliation{Universit\'{e} Paris-Saclay, CNRS/IN2P3, IJCLab, 91405 Orsay, France}

\author{D.~S.~Rosa}
\affiliation{Instituto de F\'{i}sica Te\'{o}rica, Universidade Estadual Paulista,
Rua Dr.\ Bento Teobaldo Ferraz, 271--Bloco~II, S\~{a}o Paulo, 01140-070, SP, Brazil}

\author{G.~Hupin}
\affiliation{Universit\'{e} Paris-Saclay, CNRS/IN2P3, IJCLab, 91405 Orsay, France}

\author{T.~Frederico}
\affiliation{Instituto Tecnol\'{o}gico de Aeron\'{a}utica,
Pra\c{c}a Mal.\ Eduardo Gomes, 12228-900, S\~{a}o Jos\'{e} dos Campos, SP, Brazil}

\author{M.~T.~Yamashita}
\affiliation{Instituto de F\'{i}sica Te\'{o}rica, Universidade Estadual Paulista,
Rua Dr.\ Bento Teobaldo Ferraz, 271--Bloco~II, S\~{a}o Paulo, 01140-070, SP, Brazil}

\date{\today}

\begin{abstract}
In some neutron-rich nuclei usually treated as
two-neutron halos, the core can itself be resolved into a subcore-neutron-neutron subsystem, so that four valence neutrons may be involved. To describe this situation, a four-neutron halo model in the unitary limit is developed and applied to
\(^{22}\mathrm{C}\), \(^{19}\mathrm{B}\), and \(^{14}\mathrm{Be}\), in which a
compact nuclear core is surrounded by two weakly bound spin-singlet neutron
pairs occupying different spatial shells, characterized by two independent
momentum scales, with the four-neutron wave function fully antisymmetrized.
The evolution of the halo structure with the ratio of the two scales is mapped
from the limit of well-separated scales, where the system reduces to an
effective two-neutron halo around a structured core, to the regime of
comparable scales, where the four-neutron character is fully developed. At
intermediate ratios, a window is identified in which the dimensionless
root-mean-square distances become insensitive to the scale hierarchy and the
system behaves approximately as a one-scale configuration. The calculated matter radii of \(^{22}\mathrm{C}\) and
\(^{19}\mathrm{B}\) are consistent with the most recent experimental values
and, within current uncertainties, the matter radius does not discriminate
between an effective two-neutron-halo and an explicit four-neutron-halo
description of \(^{22}\mathrm{C}\). Distinguishing the two pictures requires observables
sensitive to the shape of the halo distribution, such as ratios of higher radial moments. For \(^{14}\mathrm{Be}\), the computed
charge radius is consistent with the value derived from the measured
point-proton radius.
\end{abstract}

\maketitle

\section{Introduction}
\label{sec:intro}

Halo nuclei are characterized by a dilute cloud of weakly bound nucleons
extending far beyond the compact nuclear core~\cite{Hansen1987,AlKhalili1996}.
Because their typical size is much larger than the range of the nuclear
interaction, their low-energy structure displays universal features that are
largely insensitive to short-distance details~\cite{Zhukov1993,Bertulani2007,
Canham2008,Frederico2012,Hammer2017,Hiyama2022,Hongo2022,Naidon2023}. This
separation of scales, associated with small separation energies and large
low-energy scattering lengths, makes the unitary limit a natural reference
point for organizing the few-body dynamics of these systems.

Universal descriptions of two-neutron halos as effective three-body systems
are by now well established~\cite{Canham2008,Canham2010,Yamashita2011,
Acharya2013,Souza2016,Francisco2026}. In such approaches, the core and the two valence
neutrons form a weakly bound state whose long-distance properties are
controlled by a small number of low-energy scales, and the connection with
Efimov physics provides a framework for correlating binding energies, matter
radii, and scattering properties~\cite{Efimov1970,Naidon2017}. These
descriptions, however, treat the core as inert. For several drip-line nuclei
this assumption is questionable, because the core is itself a candidate
two-neutron halo, suggesting instead a hierarchical structure with two
distinct halo scales.

The nucleus \(^{22}\mathrm{C}\) provides a very clear example. It is usually
described as a Borromean two-neutron halo, \(^{20}\mathrm{C}+n+n\), bound
although none of its two-body subsystems is bound, with a small two-neutron
separation energy whose value is not yet known with high
precision~\cite{Tanaka2010,Yamashita2011,Acharya2013,Togano2016,Souza2016,
AME2020}. At the same time, \(^{20}\mathrm{C}\) is itself discussed as a
weakly bound \(^{18}\mathrm{C}+n+n\) halo candidate~\cite{Canham2008,
Canham2010}. This points to a two-scale picture in which an inner
\(^{20}\mathrm{C}\) halo is surrounded by two additional, more weakly bound
neutrons,
\[
    {}^{22}\mathrm{C} \;\sim\;
    \left({}^{18}\mathrm{C}+n+n\right)+n+n .
\]
The same construction applies to other drip-line systems. The two-neutron halo
\(^{19}\mathrm{B}\) is built on a \(^{17}\mathrm{B}\) core, with the unbound
\(^{18}\mathrm{B}\) subsystem providing the relevant neutron-core low-energy
scale~\cite{Hiyama2019Unit,Hiyama2022,Mazumdar2000}; regarding
\(^{17}\mathrm{B}\) schematically as a
\(^{15}\mathrm{B}+n+n\) system~\cite{Ren1990, Chu2008} leads to the analogous configuration
\(\left({}^{15}\mathrm{B}+n+n\right)+n+n\). Likewise,
\(^{14}\mathrm{Be}\) is commonly interpreted as a
\(^{12}\mathrm{Be}+n+n\) halo, and resolving \(^{12}\mathrm{Be}\) at the
cluster level into \(^{10}\mathrm{Be}+n+n\)~\cite{RomeroRedondo2008, Canham2008, Zhang2023} yields
\(\left({}^{10}\mathrm{Be}+n+n\right)+n+n\), i.e., a \(^{10}\mathrm{Be}\) core
surrounded by four weakly bound neutrons. In all three cases the question is
the same: is a description based on an inert core sufficient to capture the
spatial structure of the full system, or does resolving the internal halo
structure of the core provide additional insight?

The aim of the present work is to address this question within the universal low-energy regime using a two-scale framework that explicitly resolves the internal halo structure of the core.
The four valence neutrons are described through the
product of two three-body halo building blocks at unitarity: an inner
subsystem, corresponding to the core halo
(\(^{18}\mathrm{C}+n+n\), \(^{15}\mathrm{B}+n+n\), or
\(^{10}\mathrm{Be}+n+n\)), and an outer subsystem describing two additional
neutrons moving around the resulting inner aggregate. The two structures are
characterized by independent momentum scales \(k_A\) and \(k_B\), whose
inverses set the typical sizes of the inner and outer halos. By varying the
ratio \(k_B/k_A\), the model interpolates between a compact inner halo
surrounded by a much more extended outer halo and configurations in which the
two halos have comparable sizes. For \(^{22}\mathrm{C}\), the inner scale is
fixed by the two-neutron separation energy of \(^{20}\mathrm{C}\), while the
outer scale is varied and associated with the two-neutron separation energy of
\(^{22}\mathrm{C}\); for \(^{19}\mathrm{B}\) and \(^{14}\mathrm{Be}\), the
inner and outer scales are associated with the
\(^{15}\mathrm{B}+n+n\) and \(^{17}\mathrm{B}+n+n\) subsystems, and with the
\(^{10}\mathrm{Be}+n+n\) and \(^{12}\mathrm{Be}+n+n\) subsystems,
respectively.

The spin degrees of freedom are described by pairwise singlet states, and the
product of the two three-body building blocks is antisymmetrized over the four
neutron labels. The resulting four-neutron wave function is used to compute
one-body densities, root-mean-square distances and charge and matter radii, whose evolution we follow
as the ratio \(k_B/k_A\) is varied. In the limit of well-separated scales we
establish analytically the decoupling of the four-neutron system, with the
inner pair absorbed into a structured core and the outer pair reducing to an
effective two-neutron halo, while at intermediate ratios we identify a window
in which the dimensionless rms distances become insensitive to the scale
hierarchy and the system behaves approximately as a one-scale configuration.
A comparison with the corresponding bosonic model shows that this behavior is
governed by the Pauli principle. 

The calculated charge and matter radii are compared with
experimental values, and constraints on the outer two-neutron separation
energy are discussed. The cluster-product construction, together with the
decoupling limit, allows us to examine whether the matter radius can
distinguish between an effective two-neutron-halo and an explicit
four-neutron-halo description. For \(^{14}\mathrm{Be}\), since the charge
radii of its \(^{10}\mathrm{Be}\) and \(^{12}\mathrm{Be}\) cluster cores are
experimentally known, we additionally compute the charge radius and compare
it with the value derived from the measured point-proton radius.

This work is organized as follows. In Sec.~\ref{sec:framework} we introduce
the two-scale four-particle halo model, including the Jacobi coordinates,
the permutation structure, and the construction of the antisymmetrized wave
function. In Sec.~\ref{sec:results} we present the numerical results
in terms of dimensionless observables, analyzing the evolution of the halo
structure with the scale ratio $k_B/k_A$. In Sec.~\ref{subsec:physical} the
results are converted to physical units for \({}^{22}\mathrm{C}\),
\({}^{19}\mathrm{B}\), and \({}^{14}\mathrm{Be}\), and compared with the
experimental matter and charge radii, leading to constraints on the outer
two-neutron separation energies. Section~\ref{sec:conclusions} summarizes
the main conclusions. Details of the spin algebra are collected in Appendix~\ref{app:fermion_spin}. A complementary view of the one-body density on a logarithmic distance axis, resolving the two-cluster structure at small \(k_B/k_A\), is given in Appendix~\ref{app:log_scale}, and the decoupling of the matter radius in the scale-separated limit is derived in Appendix~\ref{app:mr}.

\section{General framework}
\label{sec:framework}

\subsection{Four-particle halo model}
\label{subsec:model}

We consider a system of four identical halo particles of mass \(m\), labeled
\(\{i,j,k,l\}\), weakly bound to a compact core of mass \(M_C\). In the nuclear
applications discussed below, these particles are neutrons, so that
\(m=m_n\). The system is organized in terms of two effective
three-body building blocks. The first, denoted by \(A\), consists of particles
\(i,j\) together with the core \(C\). The second, denoted by \(B\),
describes particles \(k,l\) moving relative to the composite inner
subsystem \(A\), whose mass is \(M_C+2m_n\). The corresponding three-body
binding momenta are denoted by \(k_A\) and \(k_B\), respectively. They define the inverse length scales associated with the inner and outer
halos, with larger values corresponding to more compact subsystems and
smaller values to more spatially extended halos.

We parametrize the two scales through the corresponding two-neutron
separation energies as
\begin{equation}
S_{2n}^{(A)}=\frac{\hbar^2 k_A^2}{m_n},
\qquad
S_{2n}^{(B)}=\frac{\hbar^2 k_B^2}{m_n},
\end{equation}
where the neutron mass is used as the reference mass. 

The dimensionless ratio, $k_B/k_A$, therefore controls the hierarchy between the outer and inner halo sizes.
The limit \(k_B/k_A\ll 1\) describes a compact inner halo surrounded by a
much more extended outer halo, whereas \(k_B/k_A\sim 1\) corresponds to
comparable inner and outer length scales.

The four-particle halo wave function is constructed as the properly
(anti)symmetrized product of two three-body wave functions at unitarity~\cite{Bulgac1976,Rosa2022,Castin2011}:
\begin{equation}
    \Psi_{4}
    =
    \mathcal{P}
    \left[
    \Psi_{ij}^{A}\otimes\Psi_{kl}^{B}
    \right],
    \label{eq:total_wf_general}
\end{equation}
where \(\mathcal{P}\) denotes the appropriate projection operator: the
antisymmetrizer \(\mathcal{A}\) for fermions or the symmetrizer
\(\mathcal{S}\) for bosons.
Each subsystem wave function is written as a product of a spatial part and an
internal state,
\begin{align}
    \Psi_{ij}^{A}
    &=
    \Phi_{A}
    \left(
    \bm r_{A_{ij}},
    \bm \rho_{A_{ij}}
    \right)
    |\chi_{A_{ij}}\rangle, \nonumber
    \\
    \Psi_{kl}^{B}
    &=
    \Phi_{B}
    \left(
    \bm r_{B_{kl}},
    \bm \rho_{B_{kl}}
    \right)
    |\chi_{B_{kl}}\rangle,
    \label{eq:subsystem_wf}
\end{align}
\noindent where the spatial parts are expressed in Jacobi coordinates.

The function \(\Phi_A\) describes the inner three-body halo and \(\Phi_B\) the outer three-body halo built on the inner aggregate. At the unitary limit, each three-body spatial wave function takes the
universal zero-range form~\cite{Bulgac1976,Rosa2022,Castin2011}. Writing the
mass-scaled Jacobi pair of subsystem \(X=A,B\) in hyperspherical
coordinates, with hyperradius and channel hyperangle
\begin{equation}
    \mathcal{R}_{X}^{2}=r_{X}^{2}+\rho_{X}^{2},
    \qquad
    \tan\alpha_{X}^{(\beta)}=\frac{r_{X}^{(\beta)}}{\rho_{X}^{(\beta)}},
    \label{eq:hyperspherical}
\end{equation}
the spatial wave function reads
\begin{align}
    \Phi_{X}\!\left(\mathcal{R}_{X},\alpha_{X}^{(\beta)}\right)
    =
    \frac{K_{\mathrm{i}s_{0}^{X}}\!\big(\sqrt{2}\,\kappa_{X}\mathcal{R}_{X}\big)}
    {\mathcal{R}_{X}^{2}}\nonumber \\
    \times \sum_{\beta=1}^{3}
    C_{X}^{(\beta)}\,
    \frac{
    \sin\!\big[\mathrm{i}s_{0}^{X}\big(\tfrac{\pi}{2}-\alpha_{X}^{(\beta)}\big)\big]
    }{
    \sin 2\alpha_{X}^{(\beta)}
    },
    \label{eq:phi_unitarity}
\end{align}
where \(K_{\mathrm{i}s_{0}^{X}}\) is the modified Bessel function of the second
kind of imaginary order and the sum runs over the three Jacobi
(spectator) channels \(\beta\), with \(C_{X}^{(\beta)}\) the corresponding channel
weights. The hyperradial scale is fixed by the three-body binding
momentum, \(\kappa_{A}=k_{A}\) for the inner subsystem and
\(\kappa_{B}=k_{B}\) for the outer one, so that \(1/\kappa_{X}\) sets the
spatial extension of each halo. The Efimov parameter \(s_{0}^{X}\) is
determined by the mass ratio of the corresponding three-body subsystem.

The states \(|\chi_{A_{ij}}\rangle\) and \(|\chi_{B_{kl}}\rangle\) contain
the internal degrees of freedom of the two pairs. For
spin-\(\frac12\) fermions these are spin-singlet states, while for spin-0
bosons the internal factor is absent.
It is important to note that the five-body wave function is constructed
from a combination of two three-body wave functions, each describing a
point-like inert core and two neutrons. Only after antisymmetrization and
the rewriting of the permuted coordinates does one obtain the wave
function for the five-body model system.

\subsection{Jacobi coordinates}
\label{subsec:jacobi}

The spatial wave functions in Eq.~(\ref{eq:subsystem_wf}) are written in
mass-scaled Jacobi coordinates. We adopt units in which \(\hbar = m_n=1\). 

For a three-body subsystem formed by two identical particles \(i,j\) and a core
\(C\) of mass \(M_C\), the \(T\)-type mass-scaled Jacobi coordinates are
\begin{align}
    \bm r_{T,ij}
    &=
    \sqrt{\frac{1}{2}}\,
    \left(\bm x_j-\bm x_i\right),
\\
    \bm\rho_{T,ij}
    &=
    \sqrt{\frac{2M_C}{M_C+2}}\,
    \left(
    \bm x_C-\frac{\bm x_i+\bm x_j}{2}
    \right),
    \label{eq:jacobi_T_scaled}
\end{align}
\noindent
where \(\bm x_i\) and \(\bm x_j\) are the position vectors of the two
identical halo particles, and \(\bm x_C\) is the position vector of the
compact core of mass \(M_C\).  Equivalently, one may use a \(Y\)-type Jacobi tree, in which one identical
particle is first paired with the core and the second identical particle is
measured relative to the center of mass of this pair. The corresponding
mass-scaled coordinates are
\begin{align}
    \bm r_{Y,ij}
    &=
    \sqrt{\frac{M_C}{M_C+1}}\,
    \left(\bm x_i-\bm x_C\right),
    \\
    \bm\rho_{Y,ij}
    &=
    \sqrt{\frac{M_C+1}{M_C+2}}\,
    \left[
    \bm x_j
    -
    \frac{M_C\bm x_C+\bm x_i}{M_C+1}
    \right].
    \label{eq:jacobi_Y_scaled}
\end{align}

\begin{figure}[t]
    \centering

    \includegraphics[
        width=0.95\columnwidth,
        trim=2.5cm 1.0cm 3.0cm 1.0cm,
        clip
    ]{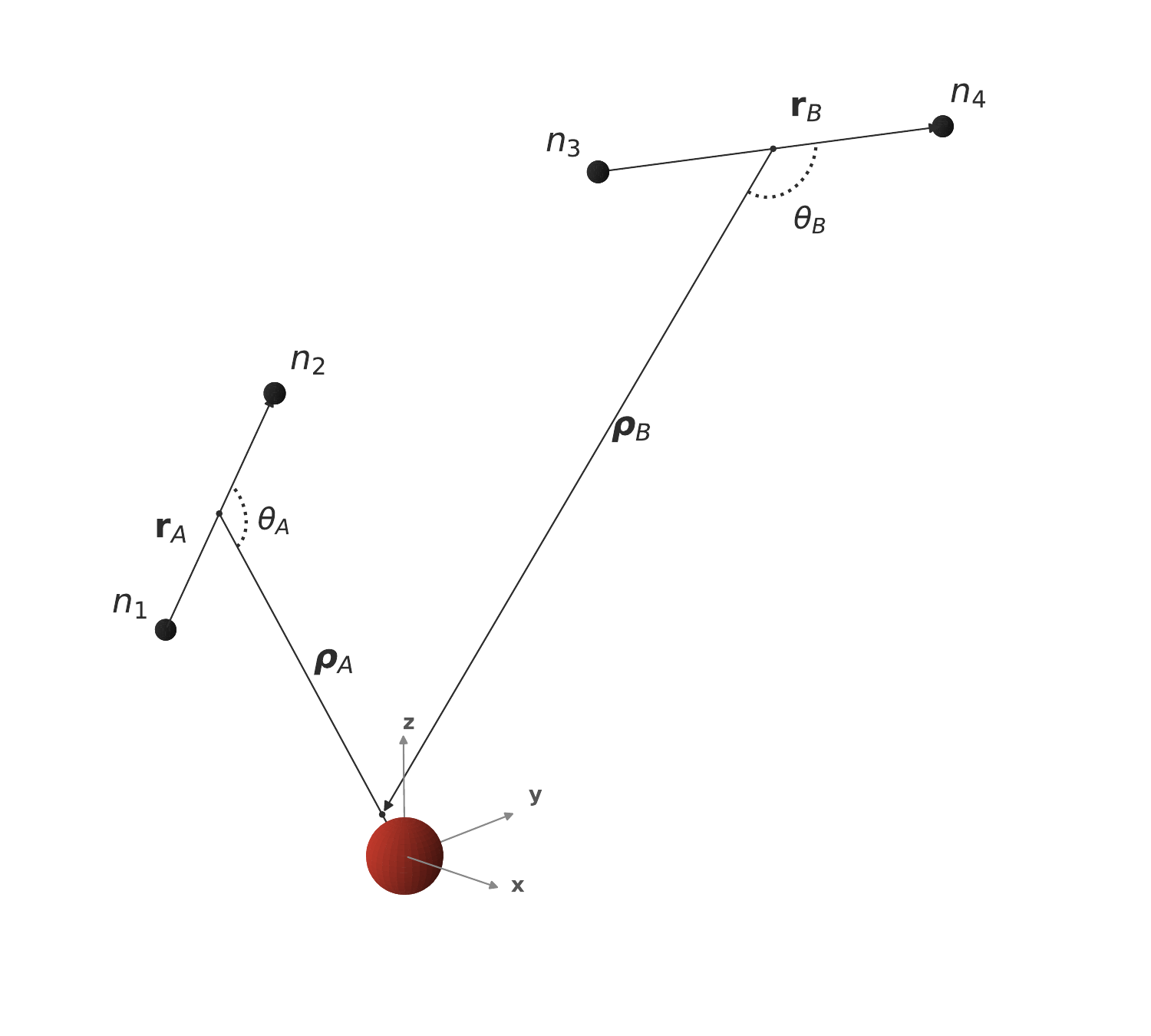}

    \vspace{-3cm}

    \includegraphics[
        width=0.95\columnwidth,
        trim=3.0cm 1.0cm 2.5cm 1.0cm,
        clip
    ]{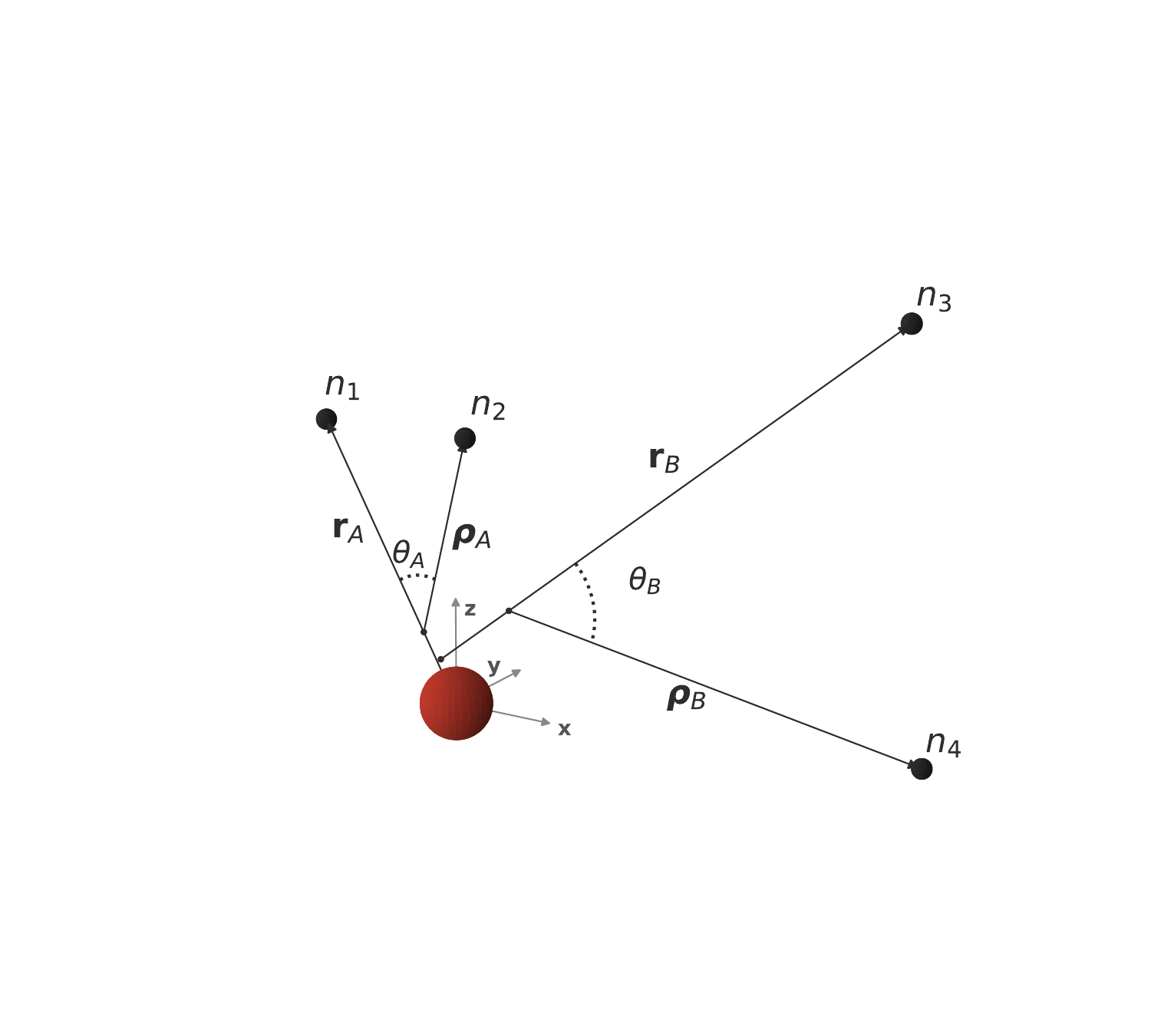}
    \vspace{-2cm}

    \caption{Jacobi coordinate systems used in the construction of the four-particle halo wave function. The \(T\)-type coordinates (upper panel) emphasize the relative motion of the two identical particles with respect to the core, while the \(Y\)-type coordinates (lower panel) describe one particle relative to the core-particle subsystem.}
    \label{fig:jacobi_coordinates}
\end{figure}

The \(T\)- and \(Y\)-type Jacobi trees used in the four-particle halo model
are depicted in Fig.~\ref{fig:jacobi_coordinates}. Both choices describe the same physical degrees of
freedom and are connected by a kinematic rotation in mass-scaled Jacobi space.
Therefore, either representation may be used depending on the observable or
numerical projection under consideration.

For the four-particle halo construction, the inner three-body subsystem \(A\)
is described by one such Jacobi pair, $(\bm r_A,\bm\rho_A)$,
while the outer subsystem \(B\) is described by a second Jacobi pair, $(\bm r_B,\bm\rho_B)$.
The effective core of subsystem \(B\) is the center of mass of the inner
subsystem \(A\),
\begin{equation}
    \bm R_A
    =
    \frac{M_C\bm x_C+\bm x_i+\bm x_j}{M_C+2}.
    \label{eq:RA_def}
\end{equation}
Thus, in the outer three-body problem, the effective core mass is
\begin{equation}
    M_A=M_C+2.
\end{equation}

In a \(T\)-type representation of the outer subsystem, the two additional
particles \(k,l\) form the first Jacobi pair and the composite subsystem \(A\)
is measured relative to their center of mass. The mass-scaled coordinates are
\begin{align}
    \bm r_{B,T;kl}
    &=
    \sqrt{\frac{1}{2}}\,
    \left(\bm x_k-\bm x_l\right),
    \\
    \bm\rho_{B,T;kl}
    &=
    \sqrt{\frac{2M_A}{M_A+2}}\,
    \left(
    \bm R_A-\frac{\bm x_k+\bm x_l}{2}
    \right).
    \label{eq:jacobi_B_T_scaled}
\end{align}

In a \(Y\)-type representation of the outer subsystem, one of the additional
particles is paired first with the composite subsystem \(A\), and the remaining
particle is measured relative to the center of mass of that pair. The
mass-scaled coordinates are
\begin{align}
    \bm r_{B,Y;k}
    &=
    \sqrt{\frac{M_A}{M_A+1}}\,
    \left(\bm x_k-\bm R_A\right),
    \\
    \bm\rho_{B,Y;l}
    &=
    \sqrt{\frac{M_A+1}{M_A+2}}\,
    \left[
    \bm x_l
    -
    \frac{M_A\bm R_A+\bm x_k}{M_A+1}
    \right].
    \label{eq:jacobi_B_Y_scaled}
\end{align}
\subsection{Permutation symmetry and coset reduction}
\label{subsec:cosets}

We start from the reference partition in which particles \(1,2\) form the
inner three-body subsystem \(A\), while particles \(3,4\) form the outer
subsystem \(B\):
\begin{equation}
    \Psi_{1234}
    =
    \Phi_{1234}\,|\xi_{1234}\rangle .
    \label{eq:Psi1234_definition}
\end{equation}
Here
\begin{equation}
    \Phi_{1234}
    =
    \Phi_A(\bm r_{A_{12}},\bm\rho_{A_{12}})
    \Phi_B(\bm r_{B_{34}},\bm\rho_{B_{34}}),
    \label{eq:Phi1234_definition}
\end{equation}
is the spatial part, while \(|\xi_{1234}\rangle\) denotes the corresponding
internal state. For spin-\(\frac12\) fermions, the identical halo particles
are coupled pairwise to spin-singlet states. Thus, for the reference
partition,
\begin{equation}
    |\xi_{1234}\rangle
    =
    |S_{12}\rangle |S_{34}\rangle ,
\end{equation}
where \(|S_{ij}\rangle\) denotes the normalized two-particle spin singlet
formed by particles \(i\) and \(j\). 
The ordering inside each spin-singlet pair is important because
the singlet is antisymmetric under exchange of its two labels, $|S_{ji}\rangle=-|S_{ij}\rangle $.

We also introduce a bosonic halo, in which the four halo particles are treated
as spinless bosons, as a reference calculation. This case is not intended as a
physical description of halo nuclei, but rather as a comparison system
constructed from the same spatial building blocks. It allows us to separate the
effects of the two-scale geometry from those associated with the
antisymmetrization of the four-neutron wave function. For spin-0 bosons, no internal spin
factor is present.

The four identical halo particles must be projected onto the appropriate
exchange-symmetry sector. The projection is obtained by summing over the
symmetric group \(S_4\):
\begin{equation}
    \mathcal P\Psi_{1234}
    =
    \sum_{P\in S_4}
    \sigma_P\,P(\Psi_{1234}),
    \label{eq:projector}
\end{equation}
where
\begin{equation}
    \sigma_P =
    \begin{cases}
        (-1)^P, & \text{fermions},\\
        +1,     & \text{bosons}.
    \end{cases}
    \label{eq:sigma_P}
\end{equation}
An overall normalization factor has been omitted, since the final wave
function is normalized after the projection.

A useful simplification follows from the symmetry under exchange inside each
preformed pair. For \(s\)-wave three-body building blocks, the spatial part is
symmetric under exchanging the two identical particles inside a pair. The
internal state then fixes the total within-pair exchange phase:
\begin{equation}
    P_{12}\Psi_{1234}
    =
    \epsilon\,\Psi_{1234},
    \qquad
    P_{34}\Psi_{1234}
    =
    \epsilon\,\Psi_{1234},
    \label{eq:within_pair}
\end{equation}
with
\begin{equation}
    \epsilon =
    \begin{cases}
        -1, & \text{spin-\(\frac12\) fermions in pair-singlet states},\\
        +1, & \text{spin-0 bosons}.
    \end{cases}
\end{equation}

Let
\begin{equation}
    H=\{e,\,(12),\,(34),\,(12)(34)\}
    \cong \mathbb Z_2\times\mathbb Z_2,
    \label{eq:subgroup_H}
\end{equation}
be the subgroup generated by the exchanges inside the two reference pairs.
For every \(h\in H\), the product of the statistical sign and the intrinsic
within-pair exchange phase satisfies
\begin{equation}
    \sigma_h\,h(\Psi_{1234})=\Psi_{1234}.
    \label{eq:H_action}
\end{equation}
Hence
\begin{equation}
    \sum_{h\in H}
    \sigma_h\,h(\Psi_{1234})
    =
    4\Psi_{1234}.
    \label{eq:H_sum}
\end{equation}

Since \(|S_4|/|H|=24/4=6\), the full permutation sum can be reduced to a sum
over six left cosets:
\begin{equation}
    S_4
    =
    \bigsqcup_{\zeta=1}^{6}g_{\zeta}H.
    \label{eq:coset_decomp}
\end{equation}
A convenient set of representatives is
\begin{equation}
    \{g_{\zeta}\}
    =
    \{e,\,(13),\,(14),\,(23),\,(24),\,(13)(24)\}.
    \label{eq:coset_reps}
\end{equation}
Therefore, after rewriting each permuted contribution in the reference set of Jacobi
coordinates, the unnormalized four-particle halo wave function can be written as
\begin{equation}
    \Psi_4(\bm r_A,\bm\rho_A,\bm r_B,\bm\rho_B)
    \equiv \Psi_4 =
    4\sum_{\zeta=1}^{6}
    \sigma_{g_{\zeta}}\,
    g_{\zeta}(\Psi_{1234}).
    \label{eq:reduced_projection}
\end{equation}
Dropping the common factor \(4\), the fermionic projected wave function is
\begin{equation}
    \Psi_4^{(F)}
    =
    \Psi_{1234}
    -\Psi_{3214}
    -\Psi_{4231}
    -\Psi_{1324}
    -\Psi_{1432}
    +\Psi_{3412}.
    \label{eq:wf_fermion_six_terms}
\end{equation}
For bosons, all statistical signs are positive:
\begin{equation}
    \Psi_4^{(B)}
    =
    \Psi_{1234}
    +\Psi_{3214}
    +\Psi_{4231}
    +\Psi_{1324}
    +\Psi_{1432}
    +\Psi_{3412}.
    \label{eq:wf_boson_six_terms}
\end{equation}

\subsection{Final form of the wave functions}
\label{subsec:grouping}

The six terms in Eqs.~(\ref{eq:wf_fermion_six_terms}) and
(\ref{eq:wf_boson_six_terms}) can be grouped in pairs. Each pair contains
spatial terms that are associated with the same two-body singlet structure after using the antisymmetry relation
of the singlet spin states. We define
the corresponding spatial combinations as
\begin{align}
    G_1
    &=
    \Phi_{1234}+\Phi_{3412},
    \nonumber\\
    G_2
    &=
    \Phi_{4231}+\Phi_{1324},
    \nonumber\\
    G_3
    &=
    \Phi_{3214}+\Phi_{1432}.
    \label{eq:spatial_grouping}
\end{align}
Therefore, the six-term fermionic expression
Eq.~(\ref{eq:wf_fermion_six_terms}) can be written as
\begin{equation}
    \Psi_4^{(F)}
    =
    \mathcal N^{(F)}
    \left[
    G_1|\xi_1\rangle
    -
    G_2|\xi_2\rangle
    +
    G_3|\xi_3\rangle
    \right],
    \label{eq:wf_fermion_grouped}
\end{equation}

\noindent where $\mathcal N^{(F)}$ is a normalization constant and

\begin{align}
|\xi_1\rangle \equiv |S_{12}\rangle |S_{34}\rangle,\nonumber \\ 
|\xi_2\rangle \equiv |S_{13}\rangle |S_{24}\rangle, \nonumber \\  
|\xi_3\rangle \equiv |S_{14}\rangle |S_{23}\rangle.
\end{align}
For spin-0 bosons there is no spin factor. Hence, the six positive terms in Eq.~(\ref{eq:wf_boson_six_terms}) reduce
directly to
\begin{equation}
    \Psi_4^{(B)}
    =
    \mathcal N^{(B)}
    \left[
    G_1+G_2+G_3
    \right],
    \label{eq:wf_boson_grouped}
\end{equation}

\noindent where $\mathcal N^{(B)}$ is the normalization constant for the bosonic wave function. 

\subsection{Probability Densities}
\label{sec:probability density}
Expanding the spin states in the product basis
\(|s_1s_2s_3s_4\rangle\), with \(s_k\in\{\uparrow,\downarrow\}\), gives
\begin{align}
    |\xi_1\rangle
    &=
    \frac{1}{2}
    \Big(
    |\!\uparrow\downarrow\uparrow\downarrow\rangle
    -
    |\!\uparrow\downarrow\downarrow\uparrow\rangle
    -
    |\!\downarrow\uparrow\uparrow\downarrow\rangle
    +
    |\!\downarrow\uparrow\downarrow\uparrow\rangle
    \Big),
    \nonumber\\[4pt]
    |\xi_2\rangle
    &=
    \frac{1}{2}
    \Big(
    |\!\uparrow\uparrow\downarrow\downarrow\rangle
    -
    |\!\uparrow\downarrow\downarrow\uparrow\rangle
    -
    |\!\downarrow\uparrow\uparrow\downarrow\rangle
    +
    |\!\downarrow\downarrow\uparrow\uparrow\rangle
    \Big),
    \nonumber\\[4pt]
    |\xi_3\rangle
    &=
    \frac{1}{2}
    \Big(
    |\!\uparrow\uparrow\downarrow\downarrow\rangle
    -
    |\!\uparrow\downarrow\uparrow\downarrow\rangle
    -
    |\!\downarrow\uparrow\downarrow\uparrow\rangle
    +
    |\!\downarrow\downarrow\uparrow\uparrow\rangle
    \Big).
    \label{eq:fermion_expansion}
\end{align}
The overlaps are obtained by identifying common product-basis states and
summing the products of their coefficients. This process is shown in detail in Appendix~\ref{app:fermion_spin}. The result is
\begin{equation}
    \mathcal S^{(F)}_{ij}
    =
    \langle \xi_i|\xi_j\rangle
    =
    \begin{pmatrix}
    1 & +\frac{1}{2} & -\frac{1}{2} \\[4pt]
    +\frac{1}{2} & 1 & +\frac{1}{2} \\[4pt]
    -\frac{1}{2} & +\frac{1}{2} & 1
    \end{pmatrix}.
    \label{eq:fermion_overlap_matrix}
\end{equation}

Using Eq.~(\ref{eq:wf_fermion_grouped}), the fermionic probability density is obtained
by tracing over spin:
\begin{equation}
    |\Psi_4^{(F)}|^2
    =
    |\mathcal N^{(F)}|^2
    \sum_{i,j=1}^{3}
    \gamma_i\gamma_j
    G_iG_j
    \mathcal S^{(F)}_{ij},
    \label{eq:fermion_prob_general}
\end{equation}
where
\begin{equation}
    \gamma_1=+1,
    \qquad
    \gamma_2=-1,
    \qquad
    \gamma_3=+1.
\end{equation}
Using  Eq.~(\ref{eq:fermion_overlap_matrix}) gives
\begin{align}
    |\Psi_4^{(F)}|^2
    =
    |\mathcal N^{(F)}|^2
    \Big[
    G_1^2+G_2^2+G_3^2 \nonumber \\
    -G_1G_2-G_1G_3-G_2G_3
    \Big].
    \label{eq:fermion_prob_density}
\end{align}
Equivalently,
\begin{equation}
    |\Psi_4^{(F)}|^2
    =
    |\mathcal N^{(F)}|^2
    \left[
    \sum_{k=1}^{3}G_k^2
    +
    \alpha^{(F)}
    \sum_{k<l}G_kG_l
    \right],
    \label{eq:fermion_alpha}
\end{equation}
with
\begin{equation}
    \alpha^{(F)}=-1.
\end{equation}
Thus, all interference terms between components with different spin
structures are destructive.

\noindent It is important to mention that, in the unitary-limit \(s\)-wave case considered here, the Faddeev components
can be chosen real. Therefore the densities are written directly in
terms of products \(G_iG_j\).

For identical spin-0 bosons, there is no internal spin degree of freedom.
Therefore, using Eq.~(\ref{eq:wf_boson_grouped}), the probability density is
\begin{align}
    |\Psi_4^{(B)}|^2
    &=
    |\mathcal N^{(B)}|^2
    (G_1+G_2+G_3)^2
    \nonumber\\
    &=
    |\mathcal N^{(B)}|^2
    \Big[
    G_1^2+G_2^2+G_3^2
    \nonumber\\ &+2G_1G_2 +2G_1G_3+2G_2G_3
    \Big].
    \label{eq:boson_prob_density}
\end{align}
Equivalently,
\begin{equation}
    |\Psi_4^{(B)}|^2
    =
    |\mathcal N^{(B)}|^2
    \left[
    \sum_{k=1}^{3}G_k^2
    +
    \alpha^{(B)}
    \sum_{k<l}G_kG_l
    \right],
    \label{eq:boson_alpha}
\end{equation}
with $\alpha^{(B)}=+2$. All interference terms are constructive.

\subsection{One-body probability densities and matter radii}
\label{subsec:rn_density}

To characterize the spatial extension of the halo, we compute the one-body density associated with the distance between one halo particle and
the total center of mass of the five-body system. In the fermionic case, this
halo particle is one of the four neutrons. We denote this distance by
\begin{equation}
    r_n
    \equiv
    \left|
    \bm x_n-\bm R_{\mathrm{CM}}
    \right|,
    \label{eq:rn_definition}
\end{equation}
where
\begin{equation}
    \bm R_{\mathrm{CM}}
    =
    \frac{
    M_C\bm x_C+\sum_{i=1}^{4}m_n\bm x_i
    }{
    M_C+4m_n
    },
    \label{eq:five_body_cm}
\end{equation}
is the total center of mass. 

The radial density \(\eta_r(r_n)\) is defined such that
\begin{equation}
    \eta_r(r_n)\,dr_n
\end{equation}
is the one-body probability of finding a selected halo particle at a distance
between \(r_n\) and \(r_n+dr_n\) from the total center of mass.

Equivalently, in terms of the full wave function, this one-body density may
be written as
\begin{equation}
    \eta_r(r_n)
    =
    \frac{1}{\mathcal N}
    \int d\Gamma\,
    |\Psi_4|^2\,
    \delta\!\left(
    r_n-\left|\bm x_n-\bm R_{\mathrm{CM}}\right|
    \right),
    \label{eq:eta_definition}
\end{equation}
where \(d\Gamma\) denotes the full internal integration measure and
\begin{equation}
    \mathcal N
    =
    \int d\Gamma\,|\Psi_4|^2,
    \label{eq:norm_definition}
\end{equation}
is the probability-density normalization. 

For the dimensionless plots in the next section, we use the scaled variable $    x = k_B r_n $.
The corresponding dimensionless one-body density \(\eta(x)\) is defined by
preserving the probability,
\begin{equation}
    \eta(x)\,dx = \eta_r(r_n)\,dr_n .
\end{equation}
Since \(dx=k_B\,dr_n\), one has
\begin{equation}
    \eta(x)
    =
    \frac{1}{k_B}\,
    \eta_r\!\left(\frac{x}{k_B}\right).
    \label{eq:eta_scaled_definition}
\end{equation}

Besides the neutron-center-of-mass density, we also compute root-mean-square (rms) distances and the matter radius
of the halo system. The matter radius measures the rms distance
of the mass distribution from the total center of mass. For a system composed
of point-like particles or clusters with masses \(m_i\) and positions
\(\bm x_i\), it is defined as
\begin{equation}
    R_m^2
    =
    \frac{1}{M_{\rm tot}}
    \sum_i m_i
    \left\langle
    \left|
    \bm x_i-\bm R_{\rm CM}
    \right|^2
    \right\rangle,
    \label{eq:general_matter_radius}
\end{equation}
where
\begin{equation}
    M_{\rm tot}=\sum_i m_i.
\end{equation}

For an effective two-neutron halo model (\(2{\rm NHM}\)), we have
\begin{equation}
    R_{m,2{\rm NHM}}^2
    =
    \frac{
    M_C\left\langle r_{c}^2\right\rangle_{2{\rm NHM}}
    +
    2\left\langle r_{n}^2\right\rangle_{2{\rm NHM}}
    }{M_C + 2},
    \label{eq:Rm_2NHM}
\end{equation}
where \(r_{c}\) is the distance between the compact core and the
total three-body center of mass, while \(r_n\) is the distance between one
halo neutron and the same center of mass. In this case, for \(^{22}{\rm C}\),
\(^{19}{\rm B}\), and \(^{14}{\rm Be}\), we have $M_C = 20$, 17, and 12,
respectively.

For the four-neutron halo model (\(4{\rm NHM}\)), the corresponding five-body matter radius is
\begin{equation}
    R_{m,4{\rm NHM}}^2
    =
    \frac{
    M_C\left\langle r_{C}^2\right\rangle_{4{\rm NHM}}
    +
    \sum_{i=1}^{4}
    \left\langle r_{n_i}^2\right\rangle_{4{\rm NHM}}
    }{M_C + 4},
    \label{eq:Rm_4NHM_general}
\end{equation}
where \(r_C\) is the distance between the compact core and the
five-body center of mass. In this case, for \(^{22}{\rm C}\),
\(^{19}{\rm B}\), and \(^{14}{\rm Be}\), we have $M_C = 18$, 15, and 10,
respectively. Since the four halo neutrons are identical, we have
\begin{equation}
    \sum_{i=1}^{4}
    \left\langle r_{n_i}^2\right\rangle_{4{\rm NHM}}
    =
    4\left\langle r_{n}^2\right\rangle_{4{\rm NHM}}.
    \label{eq:identical_neutron_rms}
\end{equation}

\begin{figure}[h]
    \centering
    \includegraphics[width=1.05\linewidth]{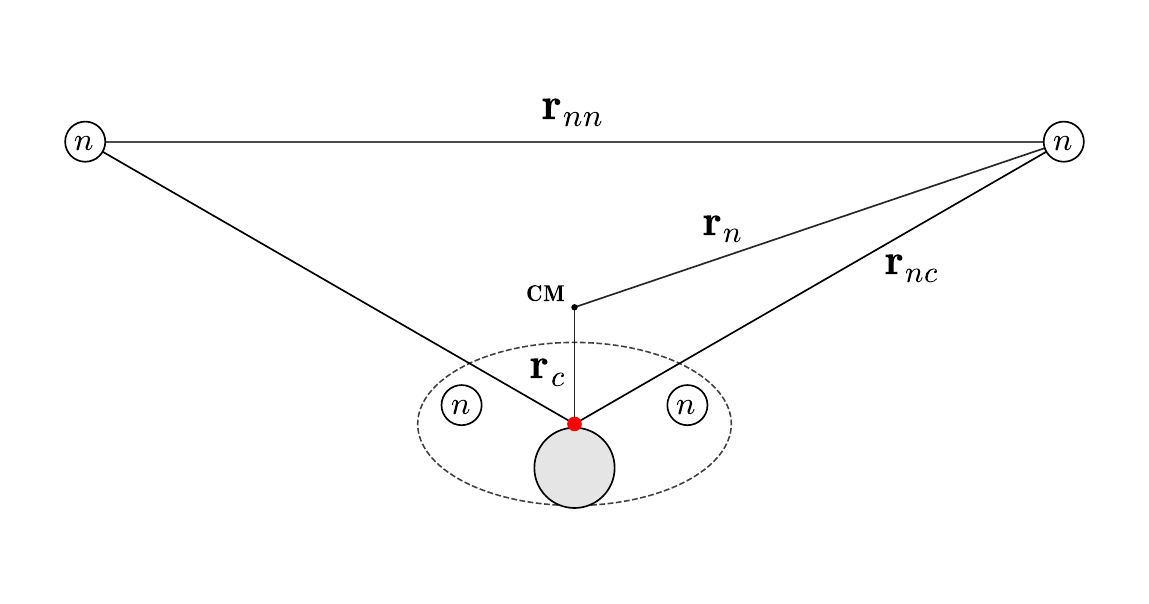}
    \centering
    \includegraphics[width=1.05\linewidth]{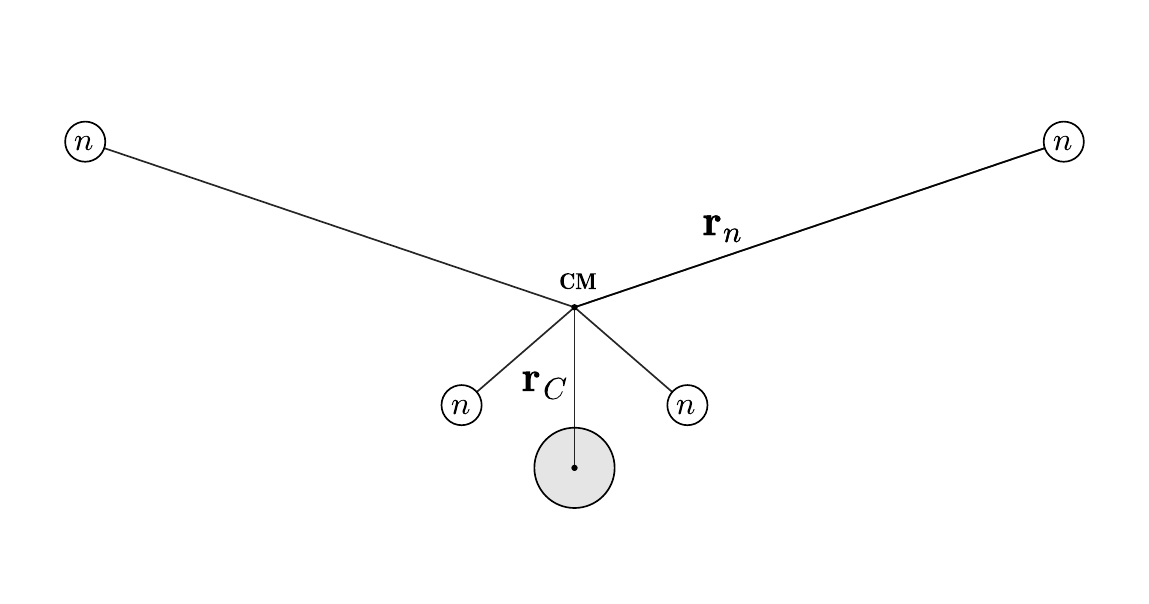}
    \caption{
Schematic representations of the two-scale four-neutron halo model. Upper panel: The inner dashed region represents the compact three-body subsystem formed by a compact core and two inner halo neutrons, which acts as an effective structured core for the outer halo. The two outer neutrons define the external halo scale, with the relevant distances $\mathbf{r}_{nn}$, $\mathbf{r}_{nc}$, $\mathbf{r}_{n}$, and $\mathbf{r}_{c}$ indicated. The position of the total center of mass is shown only schematically; in the physical system it is shifted toward the compact core because the core mass dominates over the neutron masses. The center of mass of the inner three-body subsystem is denoted by a red point. Lower panel: The distances $\mathbf{r}_{n}$ and $\mathbf{r}_{C}$ that enter the matter radius calculations are shown.
}
    \label{fig:distances}
\end{figure}

It is important to note that in the \(4{\rm NHM}\) we distinguish the distance \(r_C\) from the distance \(r_c\). The quantity \(r_C\) denotes the distance between the compact core and the
center of mass of the full few-body system under consideration, while \(r_c\) denotes the distance between the center of mass of the inner
structured core (\(^{18}{\rm C}+n+n\), \(^{15}{\rm B}+n+n\), or
\(^{10}{\rm Be}+n+n\)) and the five-body center of mass. The relevant rms distances entering these definitions are illustrated in
Fig.~\ref{fig:distances}.


\section{Dimensionless Observables}
\label{sec:results}

\begin{figure*}[!htbp]
    \centering

    \begin{minipage}{0.47\textwidth}
        \centering
        \includegraphics[width=\linewidth]{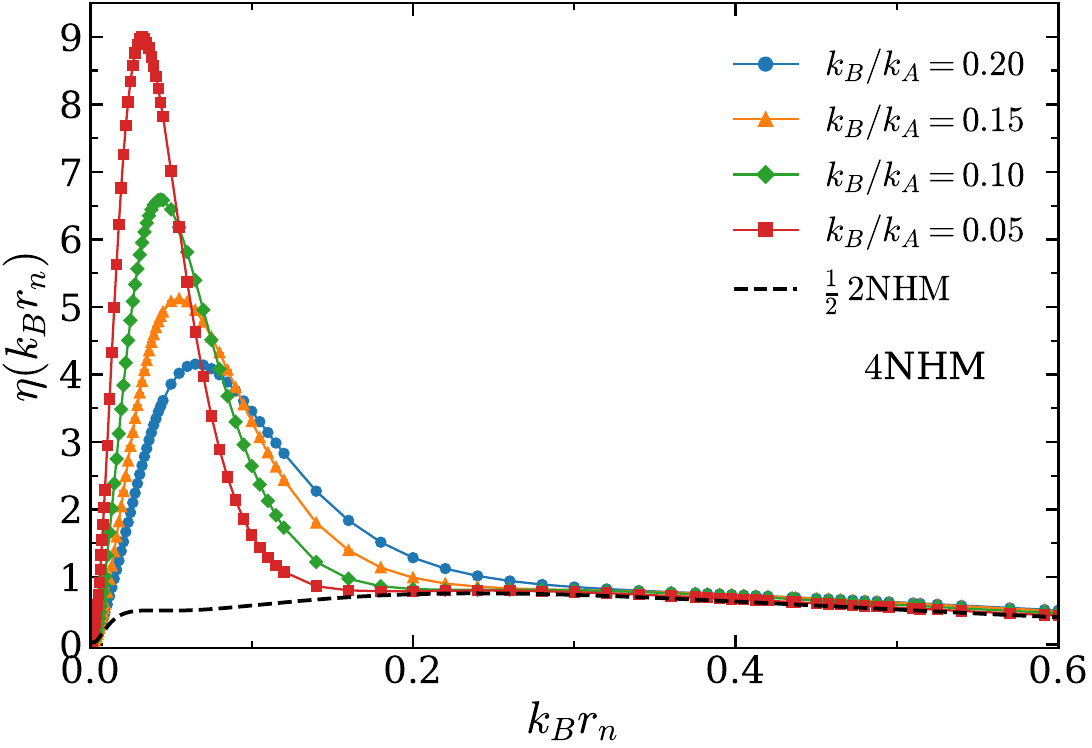}
    \end{minipage}
    \hfill
    \begin{minipage}{0.47\textwidth}
        \centering
        \includegraphics[width=\linewidth]{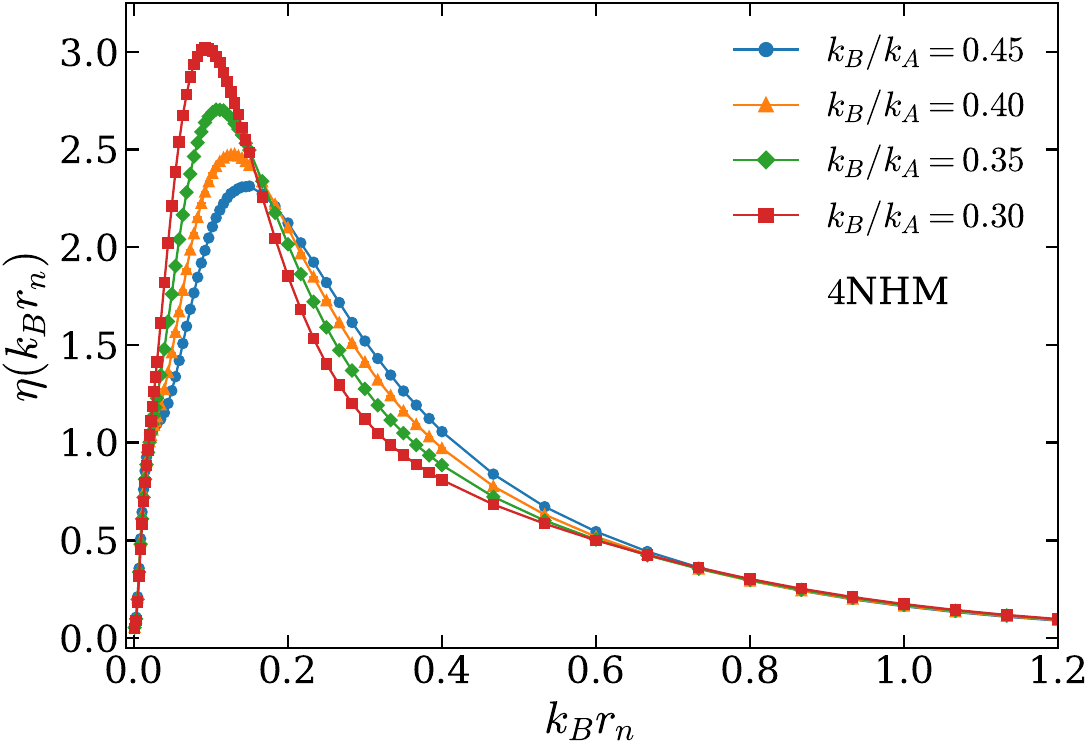}
    \end{minipage}

    \vspace{0.25cm}

    \begin{minipage}{0.47\textwidth}
        \centering
        \includegraphics[width=\linewidth]{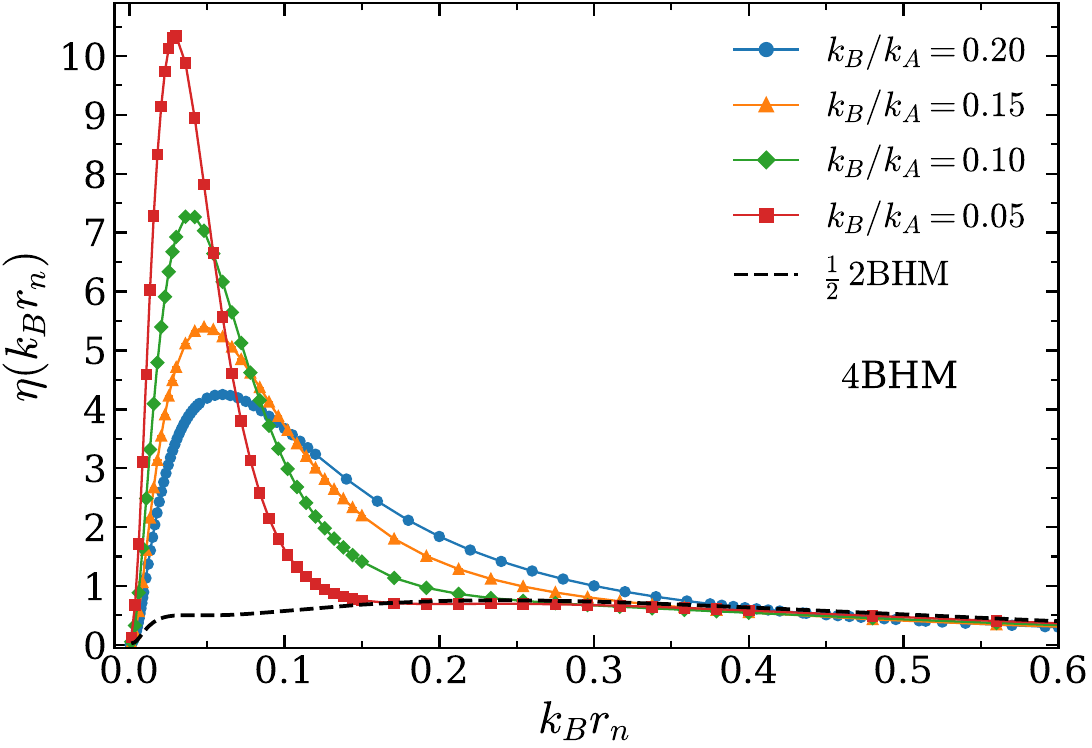}
    \end{minipage}
    \hfill
    \begin{minipage}{0.47\textwidth}
        \centering
        \includegraphics[width=\linewidth]{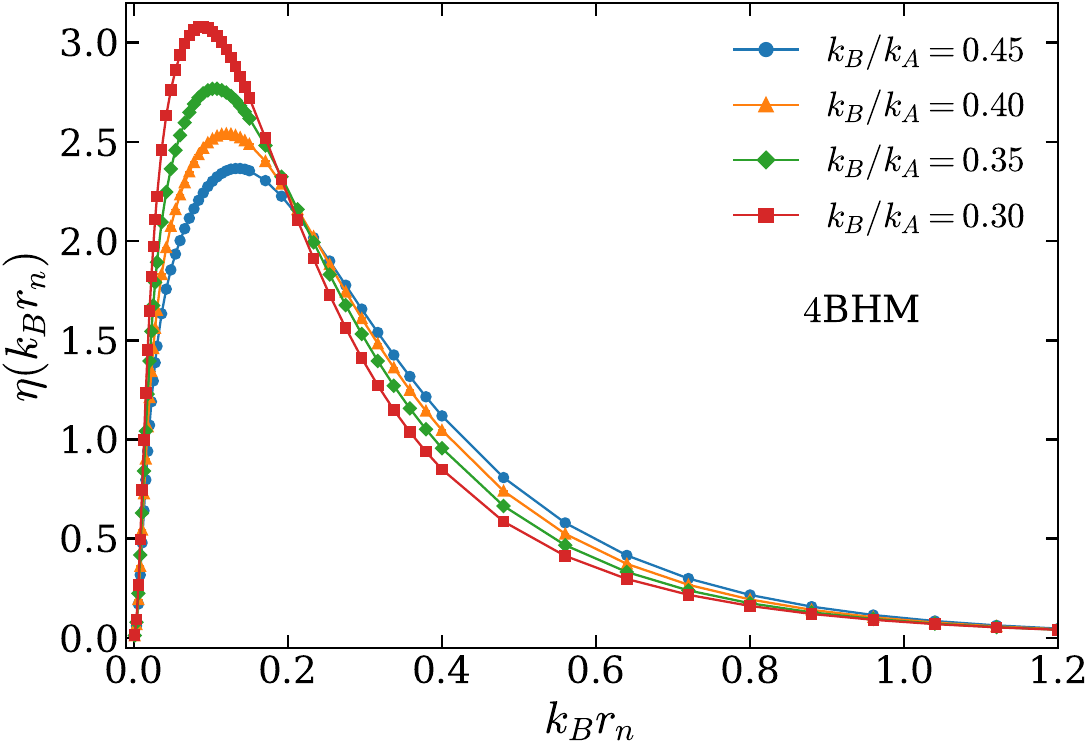}
    \end{minipage}

    \caption{
    One-body density as a function of the dimensionless distance
\(k_B r_n\) from the five-body center of mass. The different curves correspond to
different scale ratios \(k_B/k_A\), illustrating how the spatial distribution
of a single halo particle changes as the separation between the inner and outer
halo scales is varied. The dashed curve shows the effective two-particle halo
density multiplied by \(1/2\), which is the profile approached by the
long-distance region as \(k_B/k_A\to 0\). Upper panels show
the fermionic case, while lower panels show the bosonic case. To better visualize the separation between the two clusters, the small-ratio curves are also shown on a logarithmic distance axis in Fig.~\ref{fig:density_marginal_log} (Appendix~\ref{app:log_scale}).
    }
    \label{fig:density_marginal}
\end{figure*}

We now present the numerical results obtained from the four-particle halo model developed above. Throughout this section, we set $\hbar =m_{n} = 1$. The main goal is to analyze how the spatial structure of the system changes as the ratio between the outer and inner binding momenta, \(k_B/k_A\), is varied. 

The expectation values and probability densities were computed by direct
multidimensional integration of the full four-particle halo wave function. We used rotational invariance to fix the overall orientation by choosing the two Jacobi vectors of one three-body subsystem to lie in a reference plane. The orientation of the second three-body subsystem relative to the first was then parametrized by Euler angles.
The integrations were then carried out over the remaining radial variables, internal angles, and relative orientations. In this way, the geometrical correlations between the two three-body sectors and the interference terms among the grouped components \(G_i\) are retained explicitly. For the one-body densities, the same integration procedure was used, with the additional constraint that the distance of the selected particle from the center of mass was kept fixed.

As discussed in Refs.~\cite{Frederico2012,Francisco2026}, for two-neutron halo nuclei
at unitarity, the universal dimensionless rms radii,
\(\sqrt{S_{2n}\langle r_{\gamma}^2\rangle}\), approach nearly constant values
for core-to-neutron mass ratios \(A \geq 3\). Motivated by this result, and by
the small difference between the relevant mass ratios in the systems analyzed in this work, we study in this section the dimensionless one-body
densities, rms distances and matter radius using \(M_C=20\) for the \(2\)NHM and \(M_C=18\)
for the \(4\)NHM.

In Fig.~\ref{fig:density_marginal}, we present the one-body
density as a function of the dimensionless distance \(k_B r_n\) from the five-body
center of mass. The curves were calculated for several values of the ratio \(k_B/k_A\). Smaller values correspond to the regime in which the inner and outer halos are expected to be more clearly separated.
As \(k_B/k_A\) is reduced, the density peak moves toward smaller values of \(k_B r_n\) and becomes higher. Thus, when distances are measured in units of the outer length scale \(1/k_B\), the one-body distribution becomes more concentrated near the center of mass. In the four-boson halo model (4BHM), there is no Pauli restriction against the spatial overlap of the particles. The different components of the wave function can therefore contribute more coherently to configurations in which the particles occupy the same spatial region, leading to a more localized one-body density at short distances and a less extended tail at larger values of \(k_B r_n\) when compared to the 4NHM curves. 

For the moderate ratios \(k_B/k_A=0.30\)--\(0.45\), the density peak changes
only weakly, and the redistribution remains confined to short distances,
while the long-distance region remains essentially unchanged. This
confinement anticipates the approximately universal region found in the rms
distances, discussed in the following figures. In the small-ratio region,
\(k_B/k_A=0.20\)--\(0.05\), the redistribution becomes substantially more
pronounced, with the density becoming increasingly localized in the scaled
coordinate, while the long-distance region approaches the effective
two-neutron halo model (2NHM) profile, shown as the
dashed curve; the factor \(1/2\) arises because, in this limit, only half of
each neutron's density belongs to the extended component, the other half
being localized near the core. The shoulder visible in the reference curve at
short distances is a remnant of the log-periodic oscillatory behavior of the
Efimov wave function. The
fermionic and bosonic cases show the same
qualitative trend, although the 4NHM exhibits a slightly
sharper redistribution as the two length scales become more separated. A complementary view of the small-ratio curves, \(k_B/k_A = 0.05\)--\(0.20\), on a logarithmic distance axis, which resolves the two-cluster structure of the one-body density, is presented in Fig.~\ref{fig:density_marginal_log} of Appendix~\ref{app:log_scale}.

To make the role of interference explicit, we compare the integrated
four-particle halo probability density with its off-diagonal
contributions by defining the ratio

\begin{equation}
  \mathcal{R}^{(\beta)}
  =
  \frac{
    \displaystyle
    \int
    |\mathcal N^{(\beta)}|^2
    \sum_{\zeta=1}^{6}
    \big|g_{\zeta}(\Phi_{1234})\big|^{2}
    \,d\Gamma
  }{
    \displaystyle
    \int
    \big|\Psi_4^{(\beta)}\big|^{2}
    \,d\Gamma
  },
  \qquad
  \beta=F,B.
  \label{eq:density_ratio}
\end{equation}

\noindent The numerator of Eq.~(\ref{eq:density_ratio}) retains only the six diagonal contributions of the original six-term permutation expansion, before the terms are grouped into the combinations $G_i$, thereby discarding all products between distinct permuted components. Since the same normalization factor
$|\mathcal N^{(\beta)}|^2$ multiplies the diagonal and full contributions,
it cancels in the ratio.

The integrated off-diagonal contribution can be separated into two classes.
Let $W$ denote the integrated interference between the two components
belonging to the same grouping by equivalent spin structure, and let $V$
denote the integrated interference between components belonging to different
groupings. Denoting the integrated diagonal contribution by $D$, the full
integrated densities, Eqs.~(\ref{eq:fermion_alpha}) and (\ref{eq:boson_alpha}), can then be decomposed as

\begin{equation}
  \int
  \left|\Psi_4^{(\beta)}\right|^{2}
  \,d\Gamma
  =
  \left|\mathcal N^{(\beta)}\right|^{2}
  \left[D+2W+\alpha^{(\beta)}V\right].
  \label{eq:interference_decomposition}
\end{equation}

\noindent Introducing the quantities

\begin{equation}
  \bar{w}=\frac{W}{D},
  \qquad
  \bar{v}=\frac{V}{D},
\end{equation}

\noindent the ratios in Eq.~(\ref{eq:density_ratio}) become

\begin{equation}
  \mathcal{R}^{(F)}
  =
  \frac{1}{1+2\bar{w}-\bar{v}},
  \qquad
  \mathcal{R}^{(B)}
  =
  \frac{1}{1+2\bar{w}+2\bar{v}}.
  \label{eq:density_ratio_decomposition}
\end{equation}

\noindent
The contribution $2\bar{w}$ is common to both statistics, whereas the
cross-group contribution $\bar{v}$ enters with the statistics-dependent
coefficients $\alpha^{(F)}=-1$ and $\alpha^{(B)}=+2$.

\begin{figure}[h]
    \centering
    \includegraphics[width=0.95\linewidth]{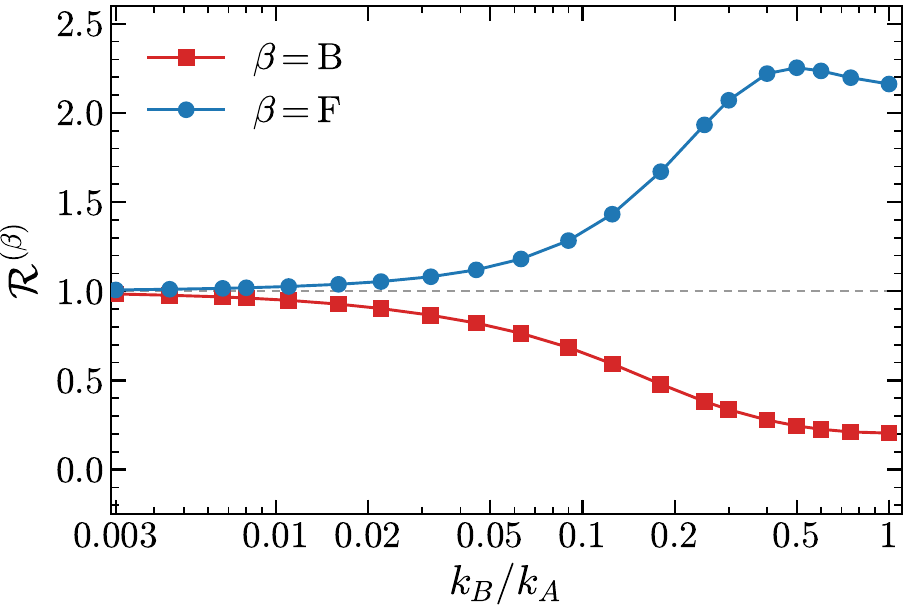}
    \centering
    \includegraphics[width=0.95\linewidth]{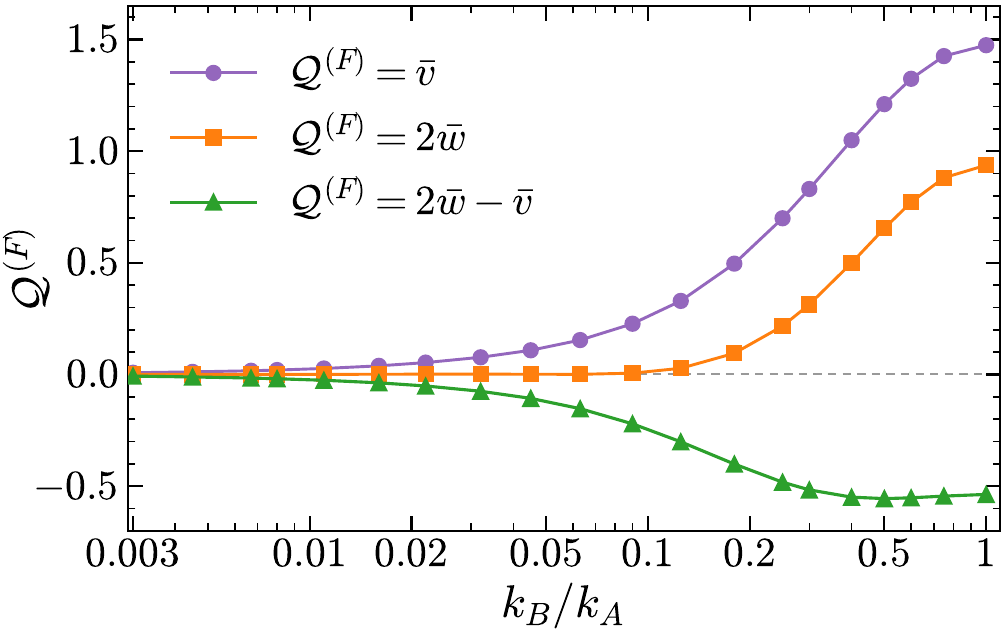}
    \caption{Upper panel: Ratio $\mathcal{R}^{(\beta)}$ of the diagonal contribution to
the full integrated four-particle halo probability density
[Eq.~(\ref{eq:density_ratio})], shown as a function of the scale ratio
$k_B/k_A$ for the fermionic ($\beta=F$) and bosonic ($\beta=B$) halos.
Lower panel: Normalized interference components
$\bar v=V/D$ and $2\bar w=2W/D$, together with the net fermionic
interference $2\bar w-\bar v$. In the scale-separated limit
$k_B/k_A\to0$, the interference contributions vanish and both ratios
approach unity, leaving only the diagonal contribution. At finite
$k_B/k_A$, the fermionic ratio exceeds unity because the destructive
contribution $-\bar v$ dominates over the constructive contribution
$2\bar w$. By contrast, the bosonic ratio falls below unity because the
two contributions combine constructively as $2\bar w+2\bar v$.}
    \label{fig:density_ratio}
\end{figure}

The upper panel of Fig.~\ref{fig:density_ratio} shows
$\mathcal{R}^{(\beta)}$ as a function of $k_B/k_A$, while the lower panel
shows the normalized interference contributions, $\mathcal{Q}^{(F)} \equiv \{\bar{v}, \ 2\bar{w}, \ 2\bar{w}-\bar{v}\}$. As the two scales separate
($k_B/k_A\to0$), the outer pair has negligible overlap with the inner
region, and both $\bar{v}$ and $\bar{w}$ vanish. Consequently, the
fermionic and bosonic ratios in the upper panel approach unity, indicating
that the integrated densities become dominated by their diagonal contributions. As $k_B/k_A$ increases, the overlap between the inner and outer regions
grows and the interference terms become increasingly important. In the
bosonic case, the two contributions reinforce one another through the
combination $2\bar{w}+2\bar{v}>0$. The full density is therefore enhanced
relative to its diagonal contribution, and $\mathcal{R}^{(B)}$ falls below
unity and decreases monotonically. The fermionic behavior is instead governed by the competition between the
constructive within-group contribution $2\bar{w}$ and the destructive
cross-group contribution $-\bar{v}$. As shown in the lower panel, over most
of the scale-ratio range $\bar{v}>2\bar{w}$, so that the net interference
$2\bar{w}-\bar{v}$ is negative. Correspondingly,
$\mathcal{R}^{(F)}$ lies above unity in the upper panel. The net
interference first becomes increasingly negative, causing
$\mathcal{R}^{(F)}$ to rise and reach a maximum at an intermediate value of
$k_B/k_A$. As the two scales become comparable, however, the growth of
$2\bar{w}$ partially compensates the destructive contribution
$-\bar{v}$. The net interference consequently becomes slightly less
negative, producing the shallow decrease of $\mathcal{R}^{(F)}$ as
$k_B/k_A$ approaches unity. The interplay between the diagonal and the different off-diagonal terms in the probability density leads to interesting behavior in the model observables, as we will see in what follows.

\begin{figure}[h]
    \centering

    \begin{minipage}{0.95\linewidth}
        \centering
        \includegraphics[width=\linewidth]{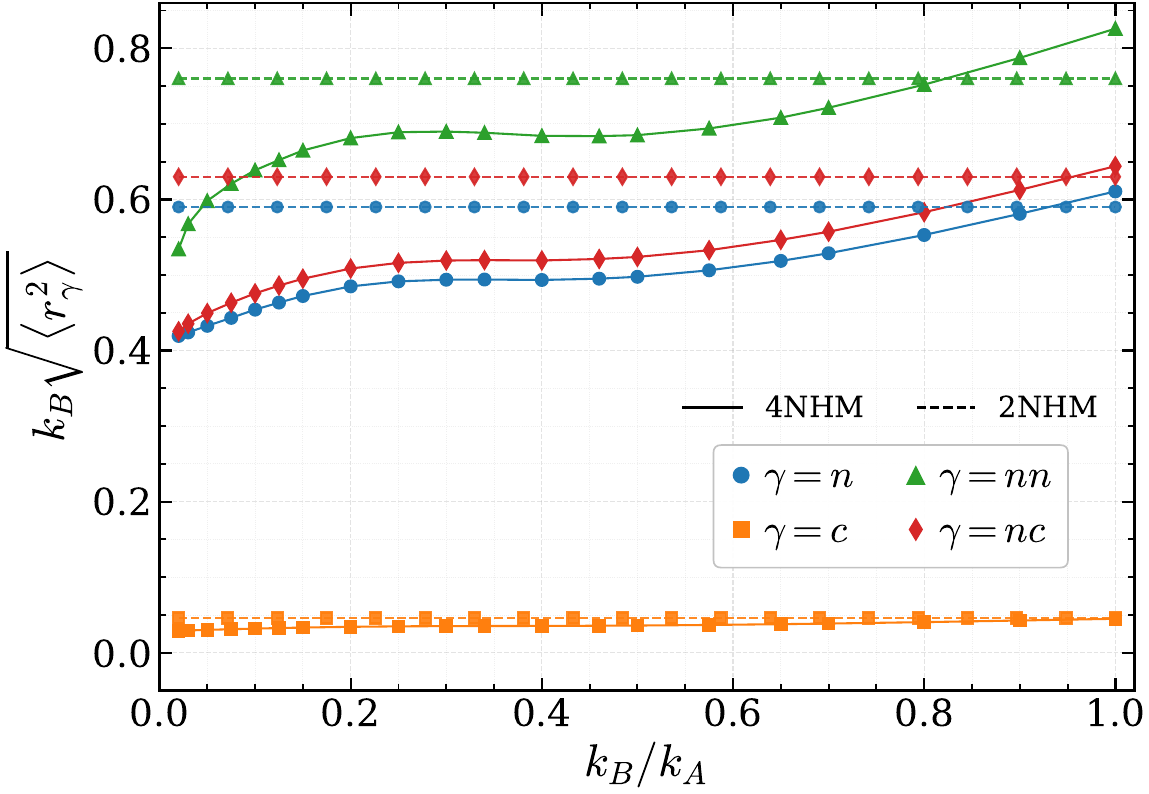}
    \end{minipage}
    \hfill
    \begin{minipage}{0.95\linewidth}
        \centering
        \includegraphics[width=\linewidth]{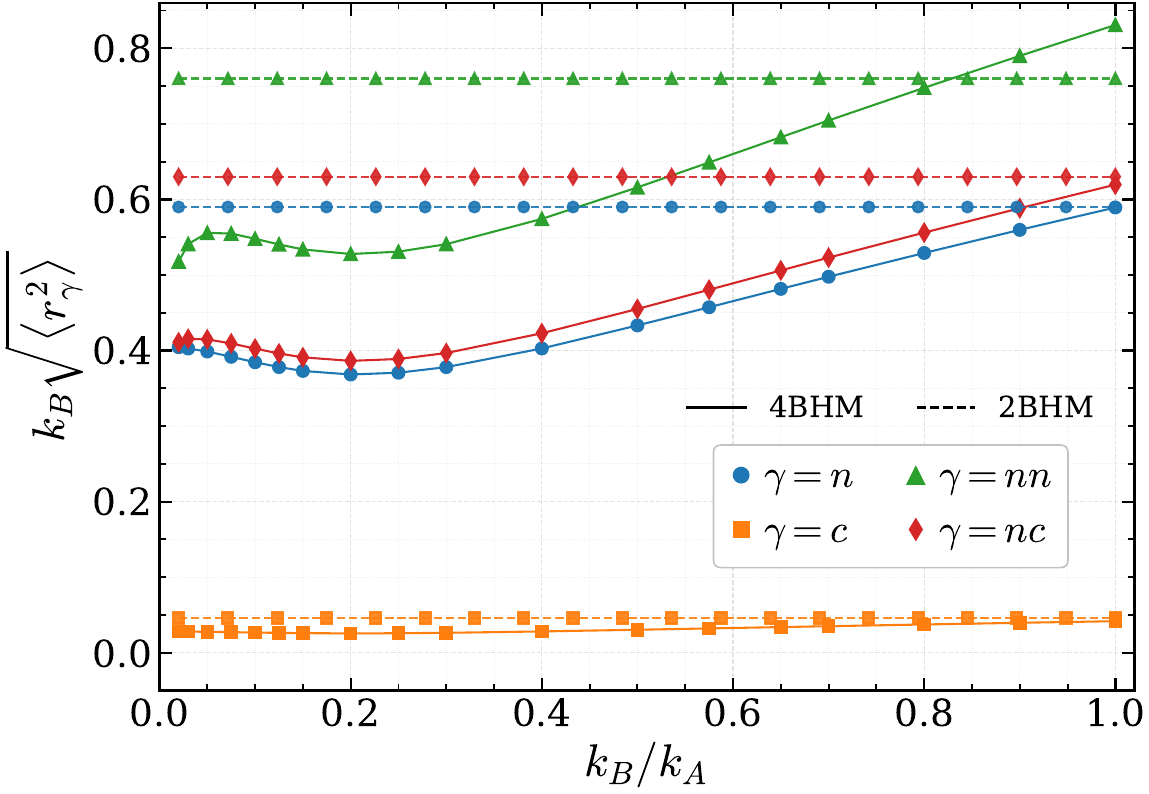}
    \end{minipage}

    \caption{Dimensionless root-mean-square distances
$k_B\sqrt{\langle r_\gamma^2\rangle}$ as functions of the scale ratio
$k_B/k_A$ for the four-particle halo model. The upper and lower panels
display the fermionic (4NHM) and bosonic (4BHM) results, respectively.
The solid curves show the four-particle halo results, while the dashed
horizontal lines indicate the corresponding effective two-particle halo
references (2NHM and 2BHM). The observables correspond to the rms
distances of a halo particle ($\gamma=n$) and of the structured core
($\gamma=c$) from the total center of mass, together with the
particle--particle ($\gamma=nn$) and particle--core ($\gamma=nc$)
relative distances.}

    \label{fig:RMS}
\end{figure}

In Fig.~\ref{fig:RMS}, we present the dimensionless rms distances
\(k_B\sqrt{\langle r_\gamma^2\rangle}\) as functions of the scale ratio
\(k_B/k_A\), comparing the 4NHM in the upper panel and the 4BHM in the lower panel with the corresponding effective three-body references, the 2NHM and the two-boson halo model (2BHM), shown as dashed lines. Since the distances are scaled by the outer momentum \(k_B\), it is
convenient to read the curves at fixed \(k_B\), so that the vertical axis then measures
the physical rms distances in units of \(1/k_B\), and decreasing \(k_B/k_A\)
corresponds to increasing \(k_A\), that is, to contracting the inner cluster
while the outer length scale is kept fixed.

The curves display three characteristic regions. For \(k_B/k_A\simeq 1\), the
inner and outer length scales are comparable and the four-particle halo rms
distances remain numerically close to the two-neutron references. This
proximity, however, does not imply a reduction of the 4NHM (4BHM) to the 2NHM
(2BHM): when the scales are comparable, the separation between inner and outer
halo particles is no longer well defined, while the four-particle halo (anti)symmetrized
structure remains explicit.
 
As \(k_B/k_A\) decreases from unity, the contraction of the inner cluster
would naively imply a monotonic decrease of all rms distances. This
expectation is indeed realized down to \(k_B/k_A\sim 0.5\), but at
intermediate ratios, roughly \(0.25\lesssim k_B/k_A\lesssim 0.5\), several
distances of the 4NHM vary only weakly. This slowly varying behavior results
from the compensation of two effects induced by the Pauli principle, whose
exclusion acts in both directions between the two pairs. On one hand, as the
inner pair contracts it blocks a progressively smaller region around the core,
and the outer neutrons can move inward, reducing the outer contribution to the
rms distances. On the other hand, the outer density now develops a sizable
amplitude in the region near the core, and the neutrons of the compact pair,
which cannot occupy states already filled by the outer ones, are partially
displaced toward larger distances; equivalently, once the two components
overlap appreciably, exchange between the identical neutrons transfers part of
the inner-pair probability to the extended configuration. The inner
contribution to the rms distances therefore increases. Over a finite window of scale
ratios these two opposite effects nearly cancel, and the dimensionless rms
distances become weakly sensitive to the detailed hierarchy between the inner
and outer binding scales. This behavior is consistent with Fig.~\ref{fig:density_marginal}, where
varying the scale ratio redistributes the density only at short distances,
with the peak following the inner scale, while the long-distance region that
dominates the rms integrals remains essentially unchanged. In this window the system behaves
approximately as an effective one-scale configuration: although the wave
function still contains the two momentum scales \(k_A\) and \(k_B\), the rms
observables do not strongly resolve their separation.
 
Finally, for \(k_B/k_A\to 0\), the inner pair collapses onto the core region and its contribution to the rms distances
vanishes. The compensation can no longer be sustained, and the fermionic
curves drop rapidly toward the decoupled limit in which only the outer pair
contributes, which is the sharp feature seen at the smallest ratios in the
upper panel. This limit can be evaluated explicitly. In this regime the two
inner neutrons are localized within a distance \(\sim 1/k_A\) of the core. The outer
neutrons, in turn, see the core and the collapsed pair as a single point-like
object and are therefore distributed as in the corresponding effective
two-neutron halo. Then, \(\langle r_n^2\rangle_{4\mathrm{NHM}}\), which is averaged over the four
neutrons, should tend to
\begin{equation}
k_B\sqrt{\langle r_n^2\rangle_{4\mathrm{NHM}}}\to
k_B\sqrt{\langle r_n^2\rangle_{2\mathrm{NHM}}}/\sqrt{2}\simeq
 0.42.
\label{eq:decoupled_limit}
\end{equation}
At the smallest scale ratios considered in Fig.~\ref{fig:RMS}, we have \(k_B\sqrt{\langle r_n^2\rangle_{4\mathrm{NHM}}}\approx 0.42\), in good agreement
with this prediction. 
 
The 4BHM departs from the naive expectation in the opposite direction. In the
absence of Pauli blocking, bosonic exchange symmetry favors the overlap
between the inner and outer pairs: as the inner cluster contracts, the outer
bosons are dragged toward the core along with it, and the rms distances
decrease below the decoupled value, producing the shallow minimum around
\(k_B/k_A\simeq 0.2\). For still smaller ratios the deeply bound inner pair
decouples, the outer distribution relaxes back toward the two-boson-halo
profile, and the rms distances rise toward the decoupled limit of
Eq.~\eqref{eq:decoupled_limit}. That the fermionic and bosonic curves meet at small ratios is
itself a consequence of the scale separation: once the inner pair no longer
overlaps with the outer one, quantum statistics between the two components
becomes irrelevant, and both the Pauli repulsion and the bosonic exchange
attraction switch off. The two models therefore approach
this common small-ratio limit from opposite sides---from above for fermions,
whose Pauli repulsion keeps the density extended, and from below for bosons,
whose exchange attraction contracts it. This contrasting behavior provides a
direct description of the role played by quantum statistics in the two-scale halo
configuration.
 
It is important to note that the individual rms distances shown in Fig.~\ref{fig:RMS} do not approach their effective two-neutron-halo counterparts in the limit $(k_B/k_A\to 0)$. In the four-particle halo model, these rms distances are averaged over all four neutrons. The compact inner pair remains close to the core and therefore reduces the average, giving rise to the factor $1/\sqrt{2}$ in Eq.~\eqref{eq:decoupled_limit}. By contrast, the effective 2NHM contains only the two outer neutrons, while the inner subsystem is incorporated into the effective core. Consequently, the observables that recover the effective 2NHM behavior in the strongly separated limit are not the rms distances, but the total matter radius, as shown in Appendix~\ref{app:mr}. This comparison is presented below.

In Fig.~\ref{fig:Matter_Radius}, we present the dimensionless matter
radius \(k_B R_m\) as a function of the scale ratio \(k_B/k_A\), obtained
from the matter radii of Eqs.~(\ref{eq:Rm_2NHM}) and
(\ref{eq:Rm_4NHM_general}). In contrast to the individual rms
distances discussed above, the matter radius contains the appropriate
center-of-mass weighting of all particles and therefore provides the
relevant global measure of the size of the full halo system. Over an
intermediate range of \(k_B/k_A\), the four-neutron results vary only
moderately, indicating that the dimensionless matter radius becomes less
sensitive to the detailed hierarchy between the inner and outer binding
scales. This behavior is consistent with the universal character of the two-scale construction seen in the previous figures. As \(k_B/k_A\) is further reduced, the
four-neutron matter radius moves toward the effective 2NHM
reference. The bosonic case shows a more pronounced nonmonotonic behavior, with the
matter radius dipping below the effective two-neutron reference at
intermediate ratios before returning to it as \(k_B/k_A\) becomes small,
whereas the fermionic case is more monotonic because antisymmetrization and
the destructive interference suppress the
inner--outer neutron overlap. This contrast reflects the role of quantum
statistics in the global size of the system. In both cases, however, the
small-ratio limit points toward the same effective two-particle halo
interpretation.
\begin{figure}[h]
    \centering
    \begin{minipage}{0.95\linewidth}
        \centering
        \includegraphics[width=\linewidth]{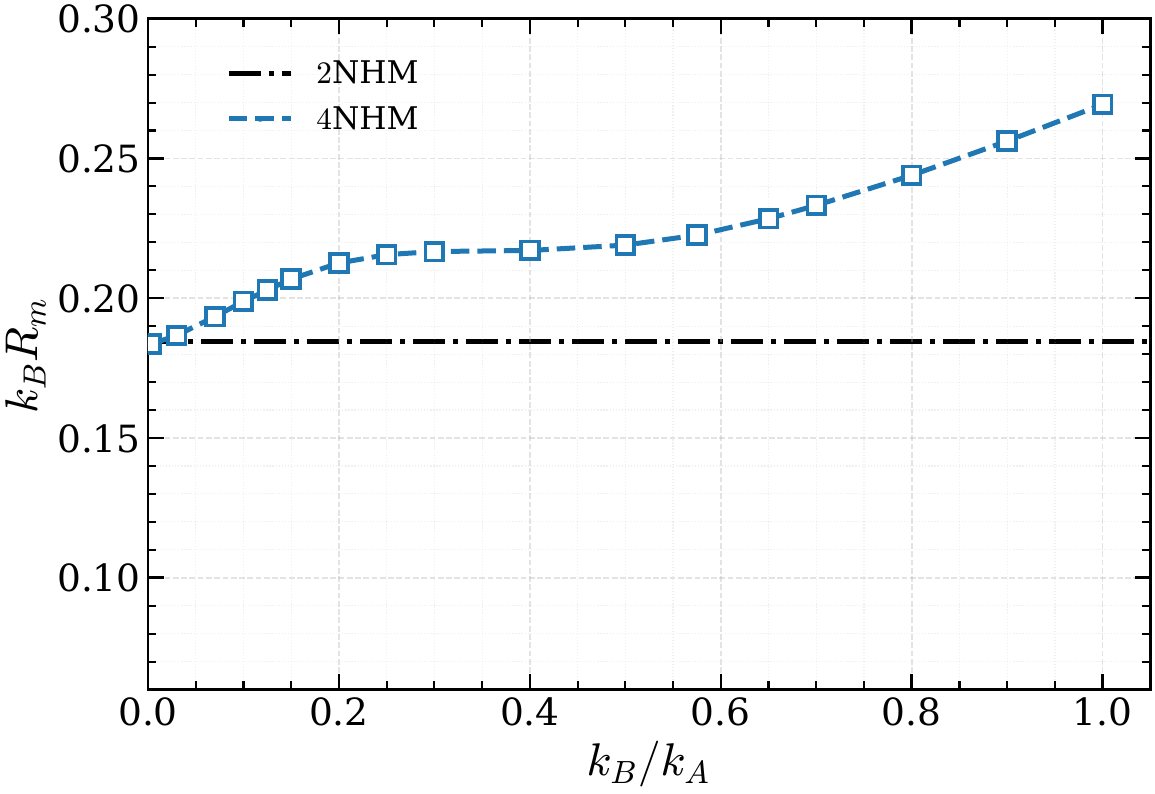}
    \end{minipage}
    \hfill
    \centering
    \begin{minipage}{0.95\linewidth}
        \centering
        \includegraphics[width=\linewidth]{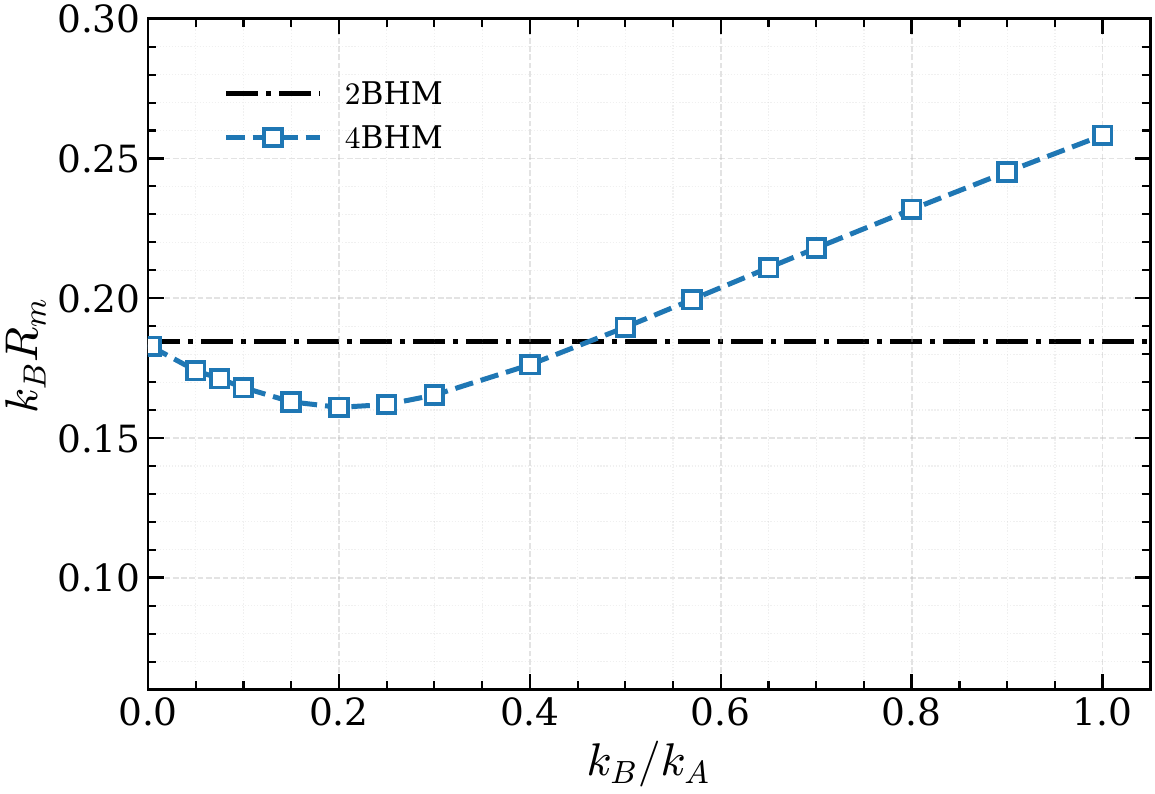}
    \end{minipage}
    \caption{Dimensionless matter radius \(k_B R_m\) as a function of the scale ratio
\(k_B/k_A\). The upper and lower panels show the four-neutron and four-boson
halo models, respectively. The dash-dotted line represents the reference
three-body result.}

    \label{fig:Matter_Radius}
\end{figure}

\section{Matter and Charge Radii}
\label{subsec:physical}

For comparison with the extracted experimental values, the dimensionless
results are converted to physical units by relating the momentum scales
\(k_A\) and \(k_B\) to the corresponding two-neutron separation
energies. The inner scale \(k_A\) is fixed by the two-neutron separation
energy of the core subsystem, while the outer scale \(k_B\) is varied
and associated with the outer two-neutron separation energy
\(S_{2n}^{(B)}\) according to
\begin{equation}
    S_{2n}^{(B)}
    =
    S_{2n}^{(A)}
    \left(\frac{k_B}{k_A}\right)^2 .
    \label{eq:S2nB}
\end{equation}
At the separation energies adopted for comparison~\cite{Togano2016,nudat3,Cook2020,nudat3,AME2020}, Eq.~\eqref{eq:S2nB} gives
\(k_B/k_A\simeq 0.40\) for \(^{22}{\rm C}\), \(0.60\) for \(^{19}{\rm B}\),
and \(0.59\) for \(^{14}{\rm Be}\), so that the three systems lie outside the
strongly separated regime, in the region where the four-neutron structure
remains fully developed. For \(^{22}{\rm C}\) and \(^{19}{\rm B}\) the
comparison is performed at the level of the matter radius, while for
\(^{14}{\rm Be}\) the measured charge radii of the beryllium cores allow us
to analyze the charge radius as well.

\subsection{\(^{22}{\rm C}\) halo system}

We first analyze \(^{22}{\rm C}\). The inner momentum scale is fixed by
the two-neutron separation energy of \(^{20}{\rm C}\), which sets the size
of the inner \(^{18}{\rm C}+n+n\) subsystem. In the numerical results
shown below we use
\(S_{2n}^{(A)}\equiv S_{2n}[^{20}{\rm C}]=3.5\,{\rm MeV}\)~\cite{nudat3}, corresponding
to \(k_A=0.2906\,{\rm fm}^{-1}\), and the outer scale is associated with
\(S_{2n}^{(B)}\equiv S_{2n}[^{22}{\rm C}]\) through Eq.~(\ref{eq:S2nB}).

Since the four-neutron construction explicitly resolves the
\(^{18}{\rm C}+n+n\) inner subsystem, the intrinsic nuclear-size
contribution to the total matter radius is taken from the matter radius of
\(^{18}{\rm C}\), rather than from an effective \(^{20}{\rm C}\) core
radius. The physical matter radius of the 4NHM
is thus obtained from
\begin{equation}
    R_{m,4{\rm NHM}}^{(\rm phys)}
    =
    \left[
    R_{m,4{\rm NHM}}^2
    +
    \frac{18}{22}\,R_{m}[^{18}{\rm C}]^2
    \right]^{1/2},
    \label{eq:Rm4NHM}
\end{equation}
with \(R_m[^{18}{\rm C}]=2.86(4)\,{\rm fm}\) from the updated
carbon-isotope analysis of Kanungo et al.~\cite{Kanungo2016}. The earlier
value \(R_m[^{18}{\rm C}]=2.82(4)\,{\rm fm}\), reported by Ozawa et
al.~\cite{Ozawa2001} and quoted in 
Ref.~\cite{Fortune2016}, produces only a small numerical variation and
does not affect the conclusions. For the 2NHM, in which \(^{22}{\rm C}\) is treated as \(^{20}{\rm C}+n+n\), the
corresponding finite-core input is the matter radius of
\(^{20}{\rm C}\),
\begin{equation}
    R_{m,2{\rm NHM}}^{(\rm phys)}
    =
    \left[
    R_{m,2{\rm NHM}}^2
    +
    \frac{20}{22}\,R_m[^{20}{\rm C}]^2
    \right]^{1/2},
    \label{eq:Rm2NHM}
\end{equation}
where we use the central value of
\(R_m[^{20}{\rm C}]=2.97^{+0.03}_{-0.05}\,{\rm fm}\)~\cite{Togano2016}.
In this way, the 4NHM and the effective 2NHM are compared at the level of
physical matter radii, each including the finite size of the corresponding
core.

The matter radius dependence on \(S_{2n}[^{22}{\rm C}]\) is especially relevant because
this quantity is not yet well determined. The large matter
radius reported by Tanaka et al.~\cite{Tanaka2010},
\(R_m[^{22}{\rm C}]=5.4(9)\,{\rm fm}\),
suggested an extremely weakly bound halo and motivated theoretical
estimates placing \(S_{2n}[^{22}{\rm C}]\) below a few hundred keV, or
even below \(0.1\)--\(0.2\,{\rm MeV}\), depending on the model
assumptions~\cite{Yamashita2011,Acharya2013,Fortune2016}. On the mass-evaluation side, the situation is similarly unsettled. The AME2020 evaluation places $S_{2n}[^{22}{\rm C}]$ at only a few tens of keV, while the direct mass measurement of Gaudefroy et al.~\cite{Gaudefroy2012} gives
\[
S_{2n}[^{22}{\rm C}]=-0.14(46)\,{\rm MeV},
\]
which is compatible with zero within its uncertainty. More recently, an uncertainty-quantified three-body analysis of
\(^{22}\mathrm{C}\) as \(^{20}\mathrm{C}+n+n\) found that comparison
with experimentally inferred matter radii favors
\(S_{2n}[^{22}\mathrm{C}]<0.35\,\mathrm{MeV}\)~\cite{McGlynn2026}. These results illustrate the present ambiguity in the two-neutron separation energy and justify treating $S_{2n}[^{22}{\rm C}]$ as a free scale parameter in our analysis, rather than fixing it to a single evaluated value. 

For the comparison with experiment, we do not use the large matter
radius reported by Tanaka et al.~\cite{Tanaka2010} as the main
constraint because of its large uncertainty. A subsequent reanalysis
of the reaction cross sections by Nagahisa and Horiuchi~\cite{Nagahisa2018}
supports a smaller matter radius, consistent with the extraction
of Togano et al.~\cite{Togano2016}. We therefore compare our results
with the more precise analysis of Togano et al.~\cite{Togano2016},
which yields
\[
R_m\!\left[^{22}\mathrm{C}\right]=3.44(8)\,\mathrm{fm},
\qquad
S_{2n}\!\left[^{22}\mathrm{C}\right]
=
0.56^{+0.27}_{-0.20}\,\mathrm{MeV}.
\]

\begin{figure}[t]
    \centering
    \includegraphics[width=0.48\textwidth]{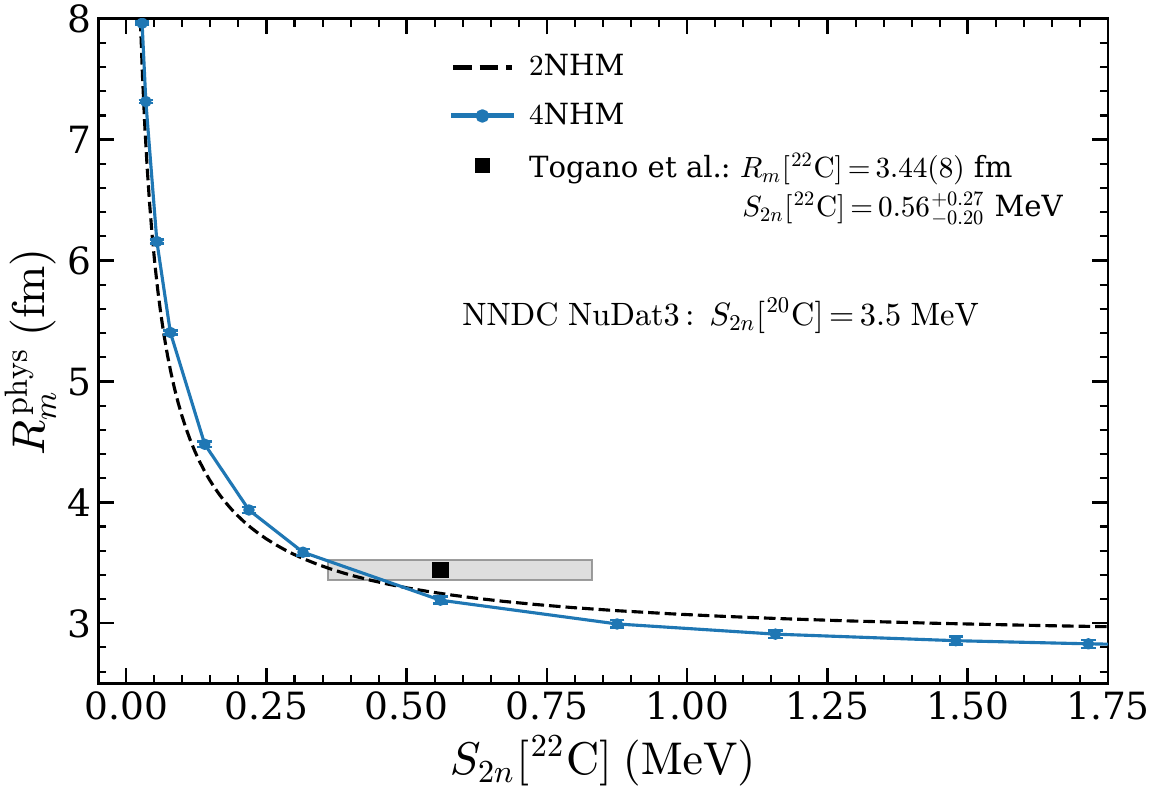}
    \caption{
Physical matter radius \(R_m^{(\mathrm{phys})}\) of \(^{22}\mathrm{C}\) as
a function of the two-neutron separation energy \(S_{2n}[^{22}{\rm C}]\)
associated with the outer halo scale. The blue points show the present
4NHM calculation, in which the \(^{18}\mathrm{C}+n+n\) inner subsystem is
explicitly resolved and the intrinsic \(^{18}\mathrm{C}\) matter radius is
included in the total radius, with the inner scale fixed by
\(S_{2n}[^{20}{\rm C}]=3.5\,\mathrm{MeV}\)~\cite{nudat3}. The dashed black curve shows
the effective 2NHM reference, in which \(^{22}\mathrm{C}\) is treated as
\(^{20}\mathrm{C}+n+n\). The black square with the shaded \(1\sigma\)
rectangle indicates the consistency region of Togano et
al.~\cite{Togano2016}, \(R_m[^{22}\mathrm{C}]=3.44(8)\,\mathrm{fm}\) and
\(S_{2n}[^{22}\mathrm{C}]=0.56^{+0.27}_{-0.20}\,\mathrm{MeV}\).
}
    \label{fig:SB_vs_Rm}
\end{figure}

Figure~\ref{fig:SB_vs_Rm} compares the calculated physical matter radius
with the experimental consistency region. The intersection of the 4NHM
curve with the experimental box selects the range of outer-halo binding
scales compatible with the extraction of Ref.~\cite{Togano2016}, providing
a phenomenological constraint on the outer halo scale within the present
two-scale framework. Within this range, the 4NHM results remain
numerically close to the effective 2NHM reference, indicating that, for
the physical scales explored here, the matter radius is mainly controlled
by the outer binding scale and is not strongly sensitive to whether the
inner \(^{18}{\rm C}+n+n\) subsystem is treated explicitly or absorbed
into an effective \(^{20}{\rm C}\) core. 

Although the matter radii of the two descriptions differ by only a few
percent in the experimentally relevant region, the corresponding neutron
distributions do not coincide, as shown in Fig.~\ref{fig:density_marginal}.
The difference lies in the shape rather than in the size of the
distribution, and can be quantified by the dimensionless ratio
\(\langle r_n^4\rangle/\langle r_n^2\rangle^2\), which is insensitive to
the overall size scale and probes the profile of the density instead. In
the one-scale 2NHM this ratio takes the universal value \(3.90\), whereas
in the two-scale 4NHM it is enhanced by the compact inner component,
ranging from \(5.0\) to \(5.2\) at the physical scale ratios
\(k_B/k_A \simeq 0.40\)--\(0.60\) of the three systems considered. As \(k_B/k_A \to 0\) the ratio approaches the decoupled value
\(2\,\langle r_n^4\rangle_{2\mathrm{NHM}}/
\langle r_n^2\rangle^2_{2\mathrm{NHM}} \simeq 7.8\) because, according to
Eq.~\eqref{eq:decoupled_limit}, both moments are halved, so the numerator
acquires a factor \(1/2\) while the squared second moment in the
denominator acquires a factor \(1/4\), yielding the overall factor of \(2\).

\subsection{\(^{19}{\rm B}\) halo system}

A similar comparison can be performed for \(^{19}\mathrm{B}\). In this
case, the two three-body components are associated with
\(^{15}\mathrm{B}+n+n\) and \(^{17}\mathrm{B}+n+n\), and the inner scale
is fixed by the two-neutron separation energy
\(S_{2n}[^{17}\mathrm{B}]=1.38\,{\rm MeV}\)~\cite{nudat3}. For the
finite-core input, analogous to Eqs.~(\ref{eq:Rm4NHM}) and
(\ref{eq:Rm2NHM}) with the appropriate mass numbers, we use the central
values of
\(R_m[^{15}\mathrm{B}]=2.59(3)\,\mathrm{fm}\)~\cite{Ozawa2001} for the
4NHM and \(R_m[^{17}\mathrm{B}]=2.99(9)\,\mathrm{fm}\)~\cite{Suzuki1999}
for the 2NHM reference.

\begin{figure}[t]
    \centering
    \includegraphics[width=0.48\textwidth]{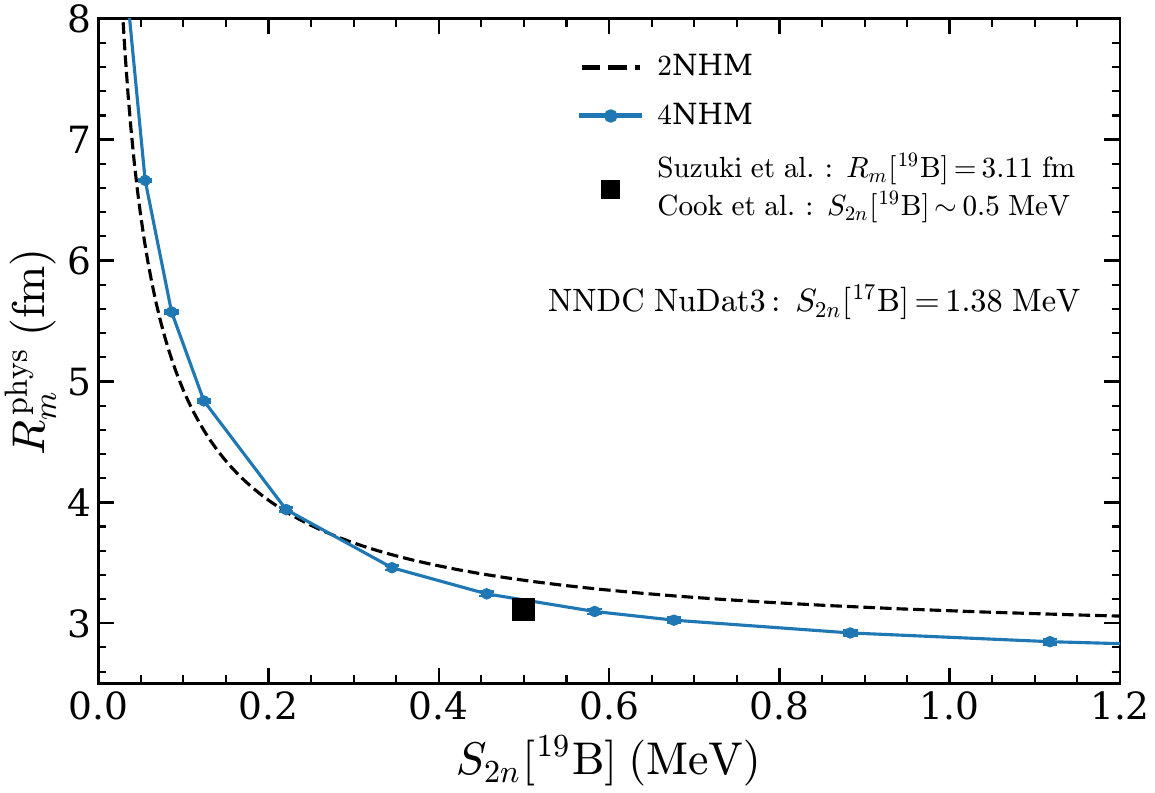}
    \caption{Physical matter radius \(R_m^{(\mathrm{phys})}\) of
\(^{19}\mathrm{B}\) as a function of the two-neutron separation energy
\(S_{2n}[^{19}{\rm B}]\) associated with the outer halo scale. The blue
points show the present 4NHM calculation, in which the
\(^{15}\mathrm{B}+n+n\) inner subsystem is explicitly resolved and the
intrinsic \(^{15}\mathrm{B}\) matter radius is included in the total
radius, with the inner scale fixed by
\(S_{2n}[^{17}{\rm B}]=1.38\,\mathrm{MeV}\). The dashed black curve shows
the effective 2NHM, in which \(^{19}\mathrm{B}\) is treated as
\(^{17}\mathrm{B}+n+n\). The black square marks the experimental matter
radius of Ref.~\cite{Suzuki1999} together with the separation energy
\(S_{2n}[^{19}\mathrm{B}]\simeq 0.5\,\mathrm{MeV}\) favored by the three-body analysis of
Ref.~\cite{Cook2020}. The uncertainties are of the order of the symbol size.}
    \label{fig:Rm_19B}
\end{figure}

The resulting physical matter radius is shown in Fig.~\ref{fig:Rm_19B} as
a function of the outer separation energy \(S_{2n}[^{19}\mathrm{B}]\),
together with the experimental matter radius of Ref.~\cite{Suzuki1999} and
the value \(S_{2n}[^{19}\mathrm{B}]\simeq 0.5\,{\rm MeV}\) obtained in
Ref.~\cite{Cook2020}. The same qualitative behavior found for
\(^{22}\mathrm{C}\) is observed: the calculated radius decreases as the
two-neutron binding increases, and the four-neutron result remains close
to the effective two-neutron results over the experimentally relevant
region. In the \(^{19}\mathrm{B}\) case, however, the separation between
the 2NHM and 4NHM curves is more visible than in the \(^{22}\mathrm{C}\)
comparison, reflecting the smaller inner binding scale and the
correspondingly larger inner halo.

As for \(^{22}\mathrm{C}\), the proximity of the two curves in the experimentally relevant region is itself a nontrivial result of the explicit four-neutron calculation: resolving the internal structure of the core could, a priori, have shifted the matter radius appreciably, but the compact inner pair is found to contribute little to the global size of the system. The matter radius thus emerges as an observable that is robust against the internal structure of the core, while the underlying neutron distributions of the four-neutron and effective two-neutron descriptions remain clearly distinguishable in their shape. This identifies observables sensitive to the shape of the halo distribution, rather than to its overall size, as the appropriate probes to discriminate between the two pictures.

\subsection{ \(^{14}{\rm Be}\) halo system}

Finally, we apply the same construction to \(^{14}\mathrm{Be}\). Here the
two three-body components are associated with \(^{10}\mathrm{Be}+n+n\)
and \(^{12}\mathrm{Be}+n+n\), and the inner scale is fixed by the
two-neutron separation energy
\(S_{2n}[^{12}\mathrm{Be}]\simeq3.67\,{\rm MeV}\)~\cite{AME2020}. The outer
scale is associated with \(S_{2n}[^{14}\mathrm{Be}]\) through
Eq.~(\ref{eq:S2nB}). For the finite-core input to the matter radius
we use the central values of
\(R_m[^{10}\mathrm{Be}]=2.30(2)\,\mathrm{fm}\)~\cite{Ozawa2001} in the
4NHM, in which the \(^{10}\mathrm{Be}+n+n\) inner subsystem is
explicitly resolved, and
\(R_m[^{12}\mathrm{Be}]=2.59(6)\,\mathrm{fm}\)~\cite{Ozawa2001} in the
effective 2NHM, in which \(^{14}\mathrm{Be}\) is treated as
\(^{12}\mathrm{Be}+n+n\).

For $^{10}\mathrm{Be}$ and $^{12}\mathrm{Be}$, experimental information on the charge radius is also available. This provides the opportunity to compare the present model with an additional observable. Direct charge-radius measurements from laser spectroscopy exist for both core candidates,
\(R_{\rm ch}[^{10}\mathrm{Be}]=2.357(18)\,\mathrm{fm}\)~\cite{Nortershauser2009}
and
\(R_{\rm ch}[^{12}\mathrm{Be}]=2.502(16)\,\mathrm{fm}\)~\cite{Krieger2012},
whereas no comparable data exist for the neutron-rich carbon and boron
cores. The charge radius of
\(^{14}\mathrm{Be}\) is built from the intrinsic charge radius
of the core and the displacement of the core relative to the total
center of mass,
\begin{equation}
    R_{\rm ch}^{(\rm phys)}
    =
    \left[
    R_{\rm ch}[{\rm core}]^2
    +
    \langle r_{\rm core\text{-}CM}^2 \rangle
    \right]^{1/2},
    \label{eq:Rch14Be}
\end{equation}
with \({\rm core}={}^{10}\mathrm{Be}\) in the 4NHM and
\({\rm core}={}^{12}\mathrm{Be}\) in the 2NHM. 

For the experimental comparison we use, in addition to the matter radius
\(R_m[^{14}\mathrm{Be}]=3.10(15)\,\mathrm{fm}\)~\cite{Suzuki1999},
extracted from interaction-cross-section measurements, and the separation energy
\(S_{2n}[^{14}\mathrm{Be}]=1.27(13)\,{\rm MeV}\)~\cite{nudat3,AME2020}, an
estimate of the \(^{14}\mathrm{Be}\) charge radius. No direct
measurement exists; however, the point-proton radius
\(R_p[^{14}\mathrm{Be}]=2.41(4)\,\mathrm{fm}\) has been determined from
charge-changing cross sections~\cite{Terashima2014}. Converting this
value with the standard relation~\cite{Kaufmann2020} between point-proton and charge radii
(including the proton and neutron mean-square charge radii and the
Darwin--Foldy term)\footnote{Explicitly,
\(R_{\rm ch}^2 = R_p^2 + \langle R_p^2\rangle_{\rm ch}
+ (N/Z)\langle r_n^2\rangle_{\rm ch} + 3\hbar^2/(4m_p^2c^2)\),
with the proton mean-square charge radius
\(\langle R_p^2\rangle_{\rm ch}\simeq 0.708\,{\rm fm}^2\), the neutron
mean-square charge radius
\(\langle r_n^2\rangle_{\rm ch}\simeq -0.116\,{\rm fm}^2\), and the
Darwin--Foldy term \(3\hbar^2/(4m_p^2c^2)\simeq 0.033\,{\rm fm}^2\).}
gives
\(R_{\rm ch}[^{14}\mathrm{Be}]\simeq 2.50(4)\,\mathrm{fm}\). This value
should be regarded as model dependent, since the underlying \(R_p\) is
extracted within a Glauber analysis; we therefore use it as a
consistency check rather than as a sharp constraint.

\begin{figure}[t]
    \centering
    \includegraphics[width=0.48\textwidth]{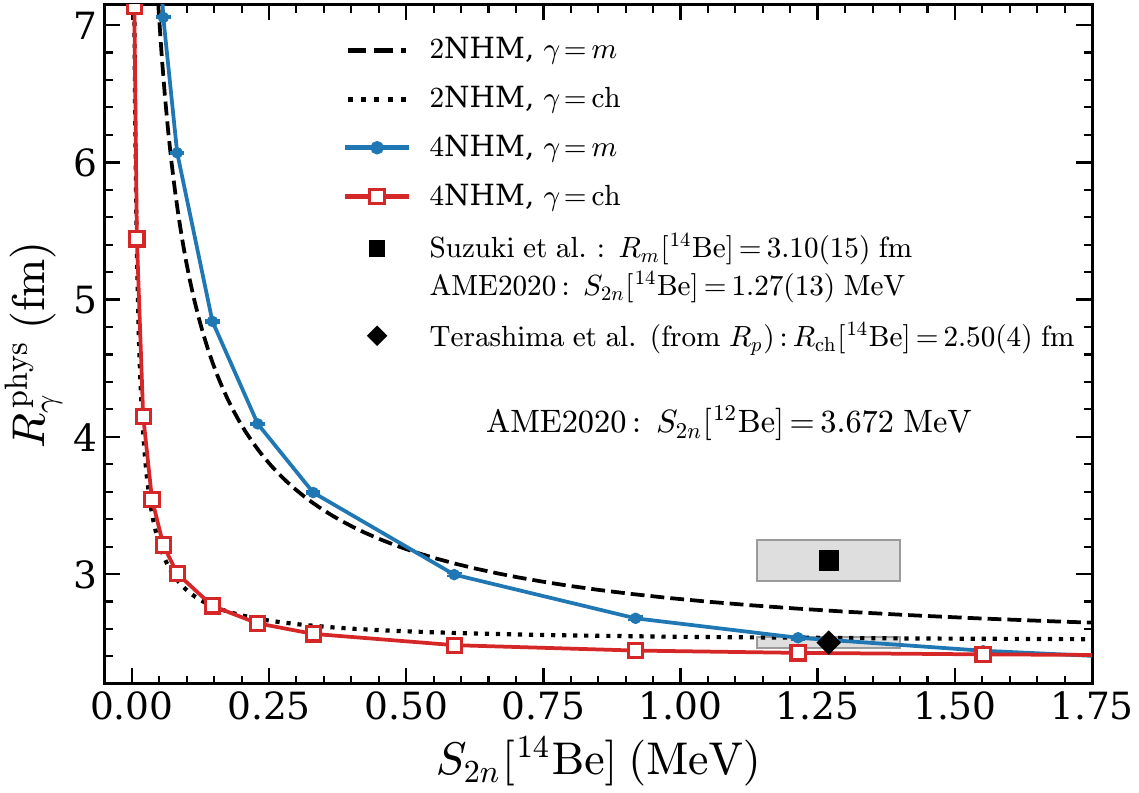}
    \caption{Physical matter and charge radii \(R_\gamma\), with
\(\gamma=m\) (\(\gamma=\mathrm{ch}\)) denoting the matter (charge)
radius, of \(^{14}\mathrm{Be}\) as functions of the two-neutron
separation energy \(S_{2n}[^{14}{\rm Be}]\) associated with the outer
halo scale. The blue circles (red open squares) show the 4NHM matter
(charge) radius, in which the \(^{10}\mathrm{Be}+n+n\) inner subsystem
is explicitly resolved and the intrinsic \(^{10}\mathrm{Be}\) matter
(charge) radius is included, with the inner scale fixed by the AME2020 value \(S_{2n}[^{12}{\rm Be}]\simeq 3.67\,{\rm MeV}\)~\cite{AME2020}. The dashed
(dotted) black curve shows the corresponding 2NHM reference, in which
\(^{14}\mathrm{Be}\) is treated as \(^{12}\mathrm{Be}+n+n\). The black
square with the shaded \(1\sigma\) rectangle marks the experimental
matter-radius constraint,
\(R_m[^{14}\mathrm{Be}]=3.10(15)\,\mathrm{fm}\)~\cite{Suzuki1999} and
\(S_{2n}[^{14}\mathrm{Be}]=1.27(13)\,\mathrm{MeV}\)~\cite{AME2020}. The
black diamond shows the charge radius
\(R_{\rm ch}[^{14}\mathrm{Be}]=2.50(4)\,\mathrm{fm}\) derived from the
point-proton radius of Ref.~\cite{Terashima2014}.}
    \label{fig:Rm_14Be}
\end{figure}

The results are shown in Fig.~\ref{fig:Rm_14Be}. As expected, both radii decrease as the two-neutron separation energy increases, while the
4NHM and 2NHM curves remain close to each other over the full range,
crossing near \(S_{2n}\simeq 0.5\,{\rm MeV}\); at the physical
separation energy the 4NHM matter radius lies slightly below the 2NHM
reference. The comparison with experiment, however, differs
qualitatively from the \(^{22}\mathrm{C}\) and \(^{19}\mathrm{B}\)
cases. Because \(S_{2n}[^{14}\mathrm{Be}]\) is comparatively well known,
the matter-radius comparison is performed at an essentially fixed outer
scale, and at \(S_{2n}=1.27(13)\,{\rm MeV}\) both descriptions
underestimate the measured matter radius: the calculated values fall
about \(0.4\)--\(0.5\,\mathrm{fm}\) below the experimental band, and the
4NHM curve would be compatible with
\(R_m[^{14}\mathrm{Be}]=3.10(15)\,\mathrm{fm}\) only for
\(S_{2n}\simeq 0.5\,{\rm MeV}\), well below the reference value. This deviation is consistent with $^{14}\mathrm{Be}$ lying farther from
the unitary regime than $^{22}\mathrm{C}$ and $^{19}\mathrm{B}$: its
outer two-neutron binding is significantly larger, and its ground state
is known to contain sizable non-$s$-wave components in the
neutron--core channels~\cite{Suzuki1999}, which are not captured by the
present universal, $s$-wave-dominated construction. The matter radius
of $^{14}\mathrm{Be}$ therefore delineates the limits of the two-scale
universal description rather than confirming its validity.

The charge radius, in contrast, is well reproduced. At the physical
separation energy both the 4NHM and 2NHM charge curves pass through the
derived value \(R_{\rm ch}[^{14}\mathrm{Be}]\simeq 2.50(4)\,\mathrm{fm}\).
For the effective 2NHM this agreement is largely built in, since the
\(^{12}\mathrm{Be}\) recoil contribution is small and the result stays
close to the intrinsic value
\(R_{\rm ch}[^{12}\mathrm{Be}]=2.502(16)\,\mathrm{fm}\). For the 4NHM,
however, the agreement is nontrivial: starting from the smaller
\(^{10}\mathrm{Be}\) charge radius, the recoil of the core against the
four halo neutrons must generate the full increase from
\(2.357(18)\,\mathrm{fm}\) to about \(2.5\,\mathrm{fm}\), and the
calculated core displacement does so at the physical scales. The charge
radius is thus less sensitive than the matter radius to the
non-universal structure of the neutron distribution, because it probes
only the center-of-mass motion of the charged core, and it remains
consistent with the data even where the matter radius signals the
breakdown of the universal description.

\section{Conclusions}
\label{sec:conclusions}

In this work, we proposed a two-scale description of a four-neutron halo
system based on the antisymmetrized product of two universal three-body halo
wave functions. The corresponding bosonic system, obtained by symmetrizing
the same two-scale product, was computed as a reference to isolate the
effects of quantum statistics. The hierarchy between the inner halo, formed
by the core and two neutrons, and the outer two-neutron halo built on this
structured subsystem is controlled by the ratio of the corresponding binding
momenta, \(k_B/k_A\), allowing us to follow the evolution of the
four-neutron halo from comparable inner and outer sizes to a compact inner
subsystem surrounded by a much more extended outer halo.

The calculated dimensionless rms distances show three characteristic
regimes. When \(k_B/k_A\simeq 1\), the inner and outer halo scales are
comparable, and the four-particle halo remains close to an effective
two-particle halo description. In the intermediate region, the scaled
spatial structure becomes weakly sensitive to variations of the scale
ratio, suggesting a universal behavior in which the two-scale system
behaves approximately as an effective one-scale configuration. Finally,
for small \(k_B/k_A\), the outer halo becomes much more extended than the
inner subsystem, which is then effectively seen by the outer particles as
a compact structured core; in this limit, the approach of the
four-particle halo matter radius toward the effective two-particle-halo
reference provides a consistency check of the two-scale
construction.

The comparison between the behavior of the fermionic and bosonic halo systems makes the role of quantum statistics in
this evolution explicit. The intermediate window of the fermionic system
originates from the compensation of two Pauli-driven effects between the
contracting inner pair and the outer neutrons, whereas in the bosonic system
the exchange attraction pulls the outer pair inward, producing a shallow
minimum instead. At small \(k_B/k_A\) the two systems reach the same
decoupled limit since statistics between
non-overlapping components becomes inoperative. These effects are largest
when the two halo scales are comparable and the overlap between inner and
outer neutrons is maximal.

We then applied the model to \(^{22}\mathrm{C}\) and \(^{19}\mathrm{B}\),
interpreted as four-neutron halos built on \(^{18}\mathrm{C}\) and
\(^{15}\mathrm{B}\) cores, with the inner scales fixed by
\(S_{2n}[^{20}\mathrm{C}]\) and \(S_{2n}[^{17}\mathrm{B}]\), respectively.
Both systems display the expected halo correlation, with larger matter radii
corresponding to smaller outer two-neutron separation energies. The compact matter
radius \(R_m[^{22}\mathrm{C}]=3.44(8)\,\mathrm{fm}\) extracted by Togano et
al.~\cite{Togano2016} intersects the calculated curve at outer bindings of a
few hundred keV, and the calculated \(^{19}\mathrm{B}\) radius is compatible
with the experimental band~\cite{Suzuki1999} in the region of the separation
energy favored by the three-body analysis of Ref.~\cite{Cook2020}. In both weakly bound systems the matter
radius thus constrains the outer halo scale, while the explicit four-neutron
description remains numerically close to the effective two-neutron reference
over the experimentally relevant region, so that the matter radius alone does
not sharply discriminate between the two pictures.

The present results trace this limited discriminating power to the observable itself rather than to any equivalence of the two descriptions: the matter radius primarily probes the overall size of the system, to which the compact inner pair — absorbed into a structureless core in the 2NHM, explicitly resolved in the 4NHM — is found to contribute very little. The underlying one-body neutron densities are, by contrast, clearly distinct: the 4NHM develops an inner peak at the scale \(1/k_A\) that has no counterpart in the 2NHM. Ratios of higher moments, which remove the dependence on the overall size and probe this profile directly, differ between the two descriptions by about \(30\%\).

The application to \(^{14}\mathrm{Be}\), with the inner scale fixed by
\(S_{2n}[^{12}\mathrm{Be}]\), plays a different role. Since
\(S_{2n}[^{14}\mathrm{Be}]=1.27(13)\,\mathrm{MeV}\)~\cite{AME2020} is
comparatively well known, the comparison is performed at an essentially
fixed outer scale and becomes a direct test of the universal description
rather than a constraint on the binding. At the physical separation
energy, both the four-neutron and the effective two-neutron descriptions
underestimate the measured matter radius by about
\(0.4\)--\(0.5\,\mathrm{fm}\); agreement would require an outer binding of
only a few hundred keV, well below the evaluated value. This deviation is
consistent with the larger outer binding of \(^{14}\mathrm{Be}\), which
places the system farther from the unitary regime, and with the known
non-\(s\)-wave components of its ground state, which are not captured by
the present \(s\)-wave-dominated construction. The matter radius of
\(^{14}\mathrm{Be}\) therefore delineates the limit of validity of the
two-scale universal picture.

The beryllium case also allowed us to analyze the charge radius, an
observable not accessible in the carbon and boron applications because
direct charge-radius measurements exist only for the beryllium cores.
The charge radius probes the
recoil of the core against the halo neutrons rather than the neutron
distribution itself. At the physical separation energy, both descriptions
reproduce the value
\(R_{\rm ch}[^{14}\mathrm{Be}]\simeq 2.50(4)\,\mathrm{fm}\) derived from
the measured point-proton radius~\cite{Terashima2014}. For the effective
two-neutron picture this agreement is largely built in, but for the
four-neutron description it is nontrivial: the core recoil must generate
the full increase from the \(^{10}\mathrm{Be}\) charge radius to the
\(^{14}\mathrm{Be}\) value, and the calculated displacement does so. The
charge radius is thus more robust than the matter radius against the
non-universal structure of the neutron distribution, remaining consistent
with the data even where the matter radius signals the breakdown of
universality.

Overall, the applications to \(^{22}\mathrm{C}\), \(^{19}\mathrm{B}\), and
\(^{14}\mathrm{Be}\) illustrate the two complementary uses of the approach:
in weakly bound systems it constrains the outer halo scale from the measured
matter radius, while in more strongly bound systems it identifies, through
the combined analysis of matter and charge radii, where the universal
description starts to fail and which observables remain protected. The
comparison with effective two-neutron-halo descriptions also shows how this
type of construction can be used to assess whether neutron-rich nuclei are
better represented as two-neutron halos or as systems with additional halo
neutrons. Since the four-neutron halo carries two momentum scales while the effective two-neutron halo carries only one, the compact inner pair — found to leave little trace in the matter radius — is naturally associated with the region of larger momenta. The differences between the two descriptions are therefore expected to be more pronounced in momentum space than in the spatial size of the system, with momentum-resolved neutron-removal measurements offering a more direct probe of the explicit four-neutron structure. We aim to explore the momentum-space structure of the two-scale halo within the present framework in future work.

In this sense, the approach may serve as a starting point for
extensions to halos with four or more weakly bound neutrons, as well as for
more refined calculations including finite scattering lengths, finite-range
corrections, more realistic core--neutron interactions, and a more detailed
treatment of spin and shell-structure effects, required in particular for a
quantitative description of \(^{14}\mathrm{Be}\).

\begin{acknowledgments}
This study was financed, in part, by the Fundação de Amparo à Pesquisa do Estado de São Paulo (FAPESP), Brazil [grant numbers 2023/13749-1 (T.F.), 2024/17816-8 (T.F. and M.T.Y.), 2025/05312-8 (T.F. and M.T.Y.), 2023/08600-9 and 2025/15267-0 (R.M.F. and T.F.)] and by the Conselho Nacional de Desenvolvimento 
Cient\'{i}fico e Tecnol\'{o}gico (CNPq) [grant numbers 306834/2022-7 (T.F.) and  
302105/2022-0 (M.T.Y.)]. This work is a part of the
project Instituto Nacional de  Ci\^{e}ncia e Tecnologia - F\'{\i}sica
Nuclear e Aplica\c{c}\~{o}es  Proc. No. CNPq 408419/2024-5. The authors thank IJCLab for hosting the collaboration.
\end{acknowledgments}

\appendix
\section{Fermionic spin-overlap computation}
\label{app:fermion_spin}

In this appendix, we detail the spin algebra required for the
4NHM. The spin part of the 4NHM wave function is built from
products of two-particle singlet states, and the antisymmetrization of the
full wave function generates all distinct pairings of the four neutrons.
The overlaps among the resulting pair-singlet states enter the norm and the
matrix elements of the antisymmetrized wave function. Below we construct
these states explicitly, expand them in the ordered product basis, and
evaluate their mutual overlaps.

\subsection{Construction of the pair-singlet states}

The singlet state for two spin-\(\frac{1}{2}\) particles \(i\) and \(j\) is
defined as
\begin{equation}
    |S_{ij}\rangle
    =
    \frac{1}{\sqrt{2}}
    \left(
    |\!\uparrow_i\downarrow_j\rangle
    -
    |\!\downarrow_i\uparrow_j\rangle
    \right).
    \label{eq:app_singlet}
\end{equation}
It satisfies
\begin{equation}
    |S_{ji}\rangle=-|S_{ij}\rangle .
    \label{eq:app_singlet_antisymmetry}
\end{equation}

The three pair-singlet states used in the fermionic four-particle halo wave
function are
\begin{align}
    |\xi_1\rangle
    &=
    |S_{12}\rangle |S_{34}\rangle,
    \nonumber\\
    |\xi_2\rangle
    &=
    |S_{13}\rangle |S_{24}\rangle,
    \nonumber\\
    |\xi_3\rangle
    &=
    |S_{14}\rangle |S_{23}\rangle .
    \label{eq:app_xi_definitions}
\end{align}
Expanding these states in the ordered product basis
\(|s_1s_2s_3s_4\rangle\), with \(s_i\in\{\uparrow,\downarrow\}\), gives
\begin{align}
    |\xi_1\rangle
    &=
    \frac{1}{2}
    \Big(
    |\!\uparrow\downarrow\uparrow\downarrow\rangle
    -
    |\!\uparrow\downarrow\downarrow\uparrow\rangle
    -
    |\!\downarrow\uparrow\uparrow\downarrow\rangle
    +
    |\!\downarrow\uparrow\downarrow\uparrow\rangle
    \Big),
    \nonumber\\[4pt]
    |\xi_2\rangle
    &=
    \frac{1}{2}
    \Big(
    |\!\uparrow\uparrow\downarrow\downarrow\rangle
    -
    |\!\uparrow\downarrow\downarrow\uparrow\rangle
    -
    |\!\downarrow\uparrow\uparrow\downarrow\rangle
    +
    |\!\downarrow\downarrow\uparrow\uparrow\rangle
    \Big),
    \nonumber\\[4pt]
    |\xi_3\rangle
    &=
    \frac{1}{2}
    \Big(
    |\!\uparrow\uparrow\downarrow\downarrow\rangle
    -
    |\!\uparrow\downarrow\uparrow\downarrow\rangle
    -
    |\!\downarrow\uparrow\downarrow\uparrow\rangle
    +
    |\!\downarrow\downarrow\uparrow\uparrow\rangle
    \Big).
    \label{eq:app_expansion}
\end{align}

\subsection{Computation of the spin overlaps}

Since the product basis is orthonormal, the overlap
\(\langle \xi_i|\xi_j\rangle\) is obtained by identifying the common product
basis states and summing the products of their coefficients.

\paragraph{Overlap \(\langle \xi_1|\xi_2\rangle\).}

The common basis vectors between \(|\xi_1\rangle\) and \(|\xi_2\rangle\) are
\[
|\!\uparrow\downarrow\downarrow\uparrow\rangle,
\qquad
|\!\downarrow\uparrow\uparrow\downarrow\rangle .
\]
Both have coefficient \(-1/2\) in \(|\xi_1\rangle\) and \(-1/2\) in
\(|\xi_2\rangle\). Therefore,
\begin{align}
    \langle \xi_1|\xi_2\rangle
    &=
    \left(-\frac{1}{2}\right)
    \left(-\frac{1}{2}\right)
    +
    \left(-\frac{1}{2}\right)
    \left(-\frac{1}{2}\right),
    \nonumber\\
    &=
    \frac{1}{4}
    +
    \frac{1}{4}
    =
    \frac{1}{2}.
    \label{eq:app_overlap_12}
\end{align}

\paragraph{Overlap \(\langle \xi_1|\xi_3\rangle\).}

The common basis vectors between \(|\xi_1\rangle\) and \(|\xi_3\rangle\) are
\[
|\!\uparrow\downarrow\uparrow\downarrow\rangle,
\qquad
|\!\downarrow\uparrow\downarrow\uparrow\rangle .
\]
In both cases, the coefficient in \(|\xi_1\rangle\) is \(+1/2\), while the
coefficient in \(|\xi_3\rangle\) is \(-1/2\). Hence
\begin{align}
    \langle \xi_1|\xi_3\rangle
    &=
    \left(+\frac{1}{2}\right)
    \left(-\frac{1}{2}\right)
    +
    \left(+\frac{1}{2}\right)
    \left(-\frac{1}{2}\right),
    \nonumber\\
    &=
    -\frac{1}{4}
    -
    \frac{1}{4}
    =
    -\frac{1}{2}.
    \label{eq:app_overlap_13}
\end{align}

\paragraph{Overlap \(\langle \xi_2|\xi_3\rangle\).}

The common basis vectors between \(|\xi_2\rangle\) and \(|\xi_3\rangle\) are
\[
|\!\uparrow\uparrow\downarrow\downarrow\rangle,
\qquad
|\!\downarrow\downarrow\uparrow\uparrow\rangle .
\]
Both have coefficient \(+1/2\) in \(|\xi_2\rangle\) and \(+1/2\) in
\(|\xi_3\rangle\). Therefore,
\begin{align}
    \langle \xi_2|\xi_3\rangle
    &=
    \left(+\frac{1}{2}\right)
    \left(+\frac{1}{2}\right)
    +
    \left(+\frac{1}{2}\right)
    \left(+\frac{1}{2}\right),
    \nonumber\\
    &=
    \frac{1}{4}
    +
    \frac{1}{4}
    =
    \frac{1}{2}.
    \label{eq:app_overlap_23}
\end{align}

Together with the normalization
\[
\langle \xi_1|\xi_1\rangle
=
\langle \xi_2|\xi_2\rangle
=
\langle \xi_3|\xi_3\rangle
=
1,
\]
these results give the spin-overlap matrix
\begin{equation}
    \mathcal S^{(F)}_{ij}
    =
    \langle \xi_i|\xi_j\rangle
    =
    \begin{pmatrix}
    1 & +\frac{1}{2} & -\frac{1}{2}
    \\[4pt]
    +\frac{1}{2} & 1 & +\frac{1}{2}
    \\[4pt]
    -\frac{1}{2} & +\frac{1}{2} & 1
    \end{pmatrix}.
    \label{eq:app_spin_overlap_matrix}
\end{equation}

\subsection{Derivation of the fermionic probability density}

The grouped fermionic wave function is
\begin{equation}
    \Psi_4^{(F)}
    =
    \mathcal N^{(F)}
    \left[
    G_1|\xi_1\rangle
    -
    G_2|\xi_2\rangle
    +
    G_3|\xi_3\rangle
    \right].
    \label{eq:app_grouped_fermion_wf}
\end{equation}
The probability density is obtained by tracing over the spin degrees of
freedom:
\begin{equation}
    |\Psi_4^{(F)}|^2
    =
    \langle \Psi_4^{(F)}|\Psi_4^{(F)}\rangle .
\end{equation}
Substituting Eq.~(\ref{eq:app_grouped_fermion_wf}) and combining the symmetric
off-diagonal terms, we obtain
\begin{align}
    |\Psi_4^{(F)}|^2
    &=
    |\mathcal N^{(F)}|^2
    \Big[
    (+G_1)(+G_1)\langle\xi_1|\xi_1\rangle
    +
    (-G_2)(-G_2)\langle\xi_2|\xi_2\rangle
    \nonumber\\
    &+
    (+G_3)(+G_3)\langle\xi_3|\xi_3\rangle
    +
    2(+G_1)(-G_2)\langle\xi_1|\xi_2\rangle
    \nonumber\\
    &+
    2(+G_1)(+G_3)\langle\xi_1|\xi_3\rangle
    +
    2(-G_2)(+G_3)\langle\xi_2|\xi_3\rangle
    \Big].
    \label{eq:app_prob_expanded}
\end{align}
Using
\begin{equation}
    \langle\xi_1|\xi_2\rangle=+\frac{1}{2},
    \qquad
    \langle\xi_1|\xi_3\rangle=-\frac{1}{2},
    \qquad
    \langle\xi_2|\xi_3\rangle=+\frac{1}{2},
\end{equation}
we find
\begin{align}
    |\Psi_4^{(F)}|^2
    &=
    |\mathcal N^{(F)}|^2
    \Big[
    G_1^2
    +
    G_2^2
    +
    G_3^2
    \nonumber\\
    &\hspace{2.0cm}
    -
    G_1G_2
    -
    G_1G_3
    -
    G_2G_3
    \Big].
    \label{eq:app_fermion_result}
\end{align}
Thus the fermionic spin structure produces destructive interference between
the three spatial pairing topologies. Terms within the same topology, such as those related by the
exchange of the inner and outer pairs, share the same spin state and
interfere without suppression.
\section{One-body density in the scale-separated limit}
\label{app:log_scale}

In this appendix we show the one-body density in the scale-separated regime, \(k_B/k_A \ll 1\), on a logarithmic scale in the dimensionless distance \(k_B r_n\), complementing the analysis of Sec.~III. 

\begin{figure}[!htbp]
    \centering
    \includegraphics[width=\linewidth]{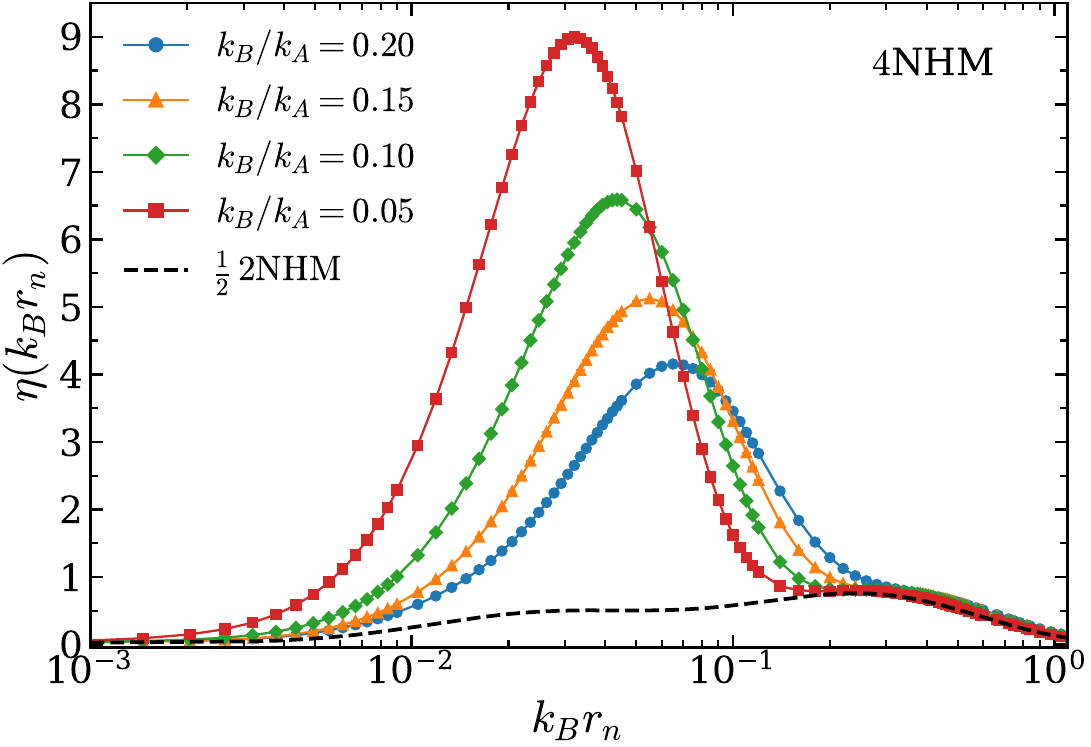}\\[4pt]
    \includegraphics[width=\linewidth]{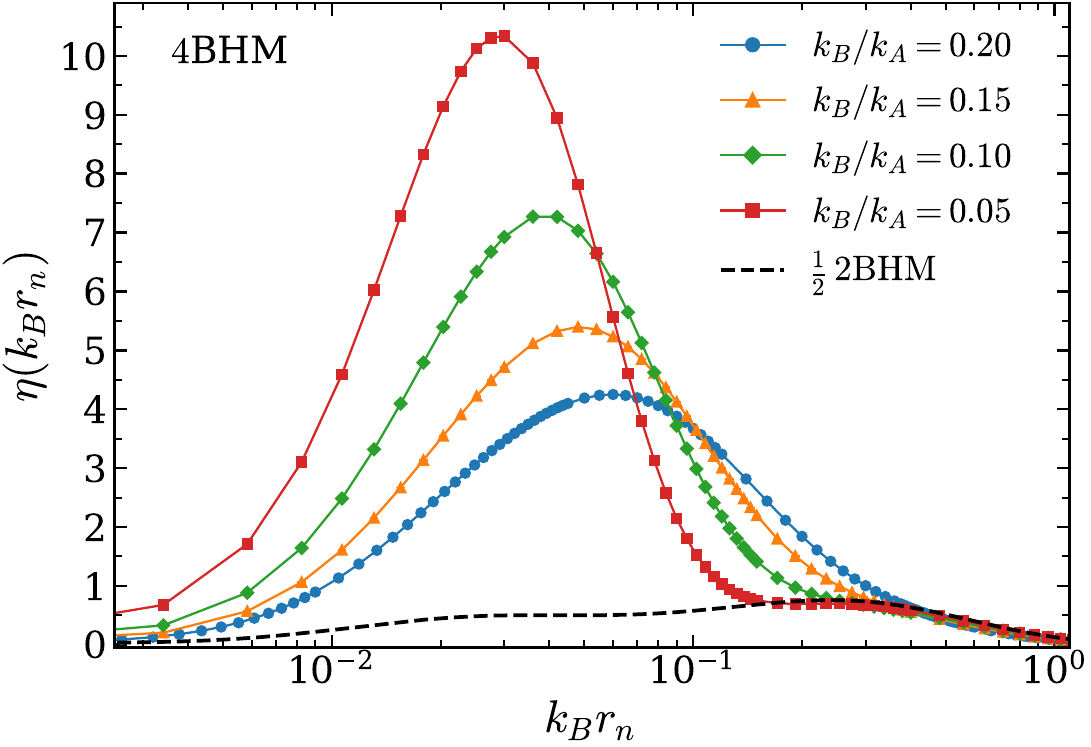}
    \caption{
    One-body density in the scale-separated regime as a function of the
    dimensionless distance \(k_B r_n\) from the five-body center of mass,
    displayed with a logarithmic \(x\) axis. The
    dashed curve shows the effective two-particle halo density multiplied by
    \(1/2\). The
    upper panel shows the fermionic case (4NHM) and the lower panel the
    bosonic case (4BHM).
    }
    \label{fig:density_marginal_log}
\end{figure}

\section{Diagonal reduction of the probability density and the matter radius
in the $k_B/k_A \to 0$ limit}
\label{app:mr}

In this appendix, we show that the diagonal part of the 4NHM mean-square
matter radius decomposes exactly into the sum of the contributions of the
inner and outer three-body subsystems. Together with the vanishing of the interference weight in this limit, established numerically in Fig.~\ref{fig:density_ratio}, this explains the approach of the full four-neutron result to the two-neutron reference observed in Fig.~\ref{fig:Matter_Radius}.

Throughout, $\bm{x}_C$ denotes the position of the compact core (mass $M_C$)
and $\bm{x}_i$ ($i=1,\dots,4$) those of the halo neutrons, with $m_{n} = 1$. The
five-body center of mass is
\begin{equation}
  \bm{R}=\frac{1}{M}\Big(M_C\,\bm{x}_C+\sum_{i=1}^{4}\bm{x}_i\Big),
  \qquad M=M_C+4 .
  \label{app:mr:cm}
\end{equation}
With $r_C=|\bm{x}_C-\bm{R}|$ and $r_{n_i}=|\bm{x}_i-\bm{R}|$, the 4NHM matter
radius is
\begin{equation}
  R^2_{m,\mathrm{4NHM}}
  =\frac{1}{M}\,
   \frac{\big\langle \Psi_4^{(\beta)}\big|\,
   \widehat{\mathcal O}\,\big|\Psi_4^{(\beta)}\big\rangle}
   {\big\langle \Psi_4^{(\beta)}\big|\Psi_4^{(\beta)}\big\rangle},
  \label{app:mr:Rmdef}
\end{equation}
\noindent where $\beta \equiv \{F,B\}$ and
\begin{equation}
\widehat{\mathcal O}=M_C\,r_C^{2}+\sum_{i=1}^{4}r_{n_i}^{2}.
  \label{app:mr:def}
\end{equation}

\subsection{Reduction to the diagonal}
\label{app:mr:diag}

In the limit $k_B/k_A\to0$, all
interference terms in the four-body densities
[Eqs.~(\ref{eq:fermion_prob_density}) and (\ref{eq:boson_prob_density})] vanish,
so that we retain only the diagonal part of the density (see Fig.~\ref{fig:density_ratio}):
\begin{equation}
    |\Psi_4^{(\beta)}|^2 \;\longrightarrow\;
    |\mathcal{N}^{(\beta)}|^2 \sum_{\zeta=1}^{6}\big|\,\Phi_\zeta\big|^{2},
    \qquad \Phi_\zeta \equiv g_\zeta(\Phi_{1234}),
  \label{app:mr:diagdens}
\end{equation}
where $\{g_\zeta\}$ is the set of six coset representatives of
Eq.~(\ref{eq:coset_reps}) and $\Phi_{1234}$ is the reference wave-function product. Inserting
Eq.~(\ref{app:mr:diagdens}) into Eq.~(\ref{app:mr:Rmdef}), the common
normalization cancels and the diagonal matter radius reads
\begin{equation}
  R^2_{m,\mathrm{4NHM}}\big|_{\mathrm{diag}}
  =\frac{1}{M}\,
   \frac{\displaystyle\sum_{\zeta=1}^{6}\big\langle \Phi_\zeta\big|\,\widehat{\mathcal O}\,\big|\Phi_\zeta\big\rangle}
        {\displaystyle\sum_{\zeta=1}^{6}\big\langle \Phi_\zeta\big|\Phi_\zeta\big\rangle}.
  \label{app:mr:diagsum}
\end{equation}

\subsection{Cluster decomposition of the diagonal terms}
\label{app:mr:steiner}

Each diagonal component $\Phi_\zeta=g_\zeta(\Phi_{1234})$ is a product in which two
neutrons are bound in the inner subsystem and two in the outer one, the
assignment being fixed by $g_\zeta$. For a general component, $\Phi_{\zeta}$, let $A_\zeta$ be the compact
subsystem consisting of the core and its two inner neutrons, with total mass
$M_A=M_C+2$ and center of mass $\bm{R}_{A_\zeta}$. Applying the parallel-axis
(Steiner) theorem to the constituents of $A_\zeta$ recasts the operator
$\widehat{\mathcal O}$ as
\begin{equation}
  \widehat{\mathcal O}=I_{A_\zeta}+I_{B_\zeta},
  \label{app:mr:steiner_a}
\end{equation}
an exact identity for every configuration, where $I_{A_\zeta}$ is the internal
moment of $A_\zeta$ and $I_{B_\zeta}$ that of the outer three-body system in which $A_\zeta$
enters as a point of mass $M_A$ at $\bm{R}_{A_\zeta}$:
\begin{align}
  I_{A_{\zeta}} &= M_C(\bm{x}_C-\bm{R}_{A_\zeta})^2+(\bm{x}_i-\bm{R}_{A_\zeta})^2+(\bm{x}_j-\bm{R}_{A_\zeta})^2 ,
  \label{app:mr:IA}\\[2pt]
  I_{{B}_{\zeta}} &= M_A(\bm{R}_{A_\zeta}-\bm{R})^2+(\bm{x}_k-\bm{R})^2+(\bm{x}_l-\bm{R})^2 .
  \label{app:mr:IB}
\end{align}
The center of mass of the outer system coincides with the
five-body one, since $M_A\bm{R}_{A_\zeta}+\bm{x}_{k}+\bm{x}_{l}=M\bm{R}$.

Substituting Eq.~(\ref{app:mr:steiner_a}) into the numerator of
Eq.~(\ref{app:mr:diagsum}) splits it into an inner and an outer sum,
\begin{equation}
  \sum_{\zeta=1}^{6}\big\langle\Phi_\zeta\big|\widehat{\mathcal O}\big|\Phi_\zeta\big\rangle
  =\sum_{\zeta=1}^{6}\big\langle\Phi_\zeta\big|I_{A_\zeta}\big|\Phi_\zeta\big\rangle
  +\sum_{\zeta=1}^{6}\big\langle\Phi_\zeta\big|I_{B_\zeta}\big|\Phi_\zeta\big\rangle .
  \label{app:mr:splitsum}
\end{equation}

The integrations in
$\langle\Phi_\zeta|I_{A_\zeta}|\Phi_\zeta\rangle$ and $\langle\Phi_\zeta|I_{B_\zeta}|\Phi_\zeta\rangle$ run over the complete space of the four
neutron coordinates, so that the change of integration variables induced by $g_\zeta$
is a mere relabeling of dummy variables and leaves the integral unchanged. 
Each sum therefore reduces to six identical contributions, and the common factor
of six cancels between numerator and denominator of Eq.~(\ref{app:mr:diagsum}). Keeping $g_\zeta = e$ gives

\begin{equation}
  R^2_{m,\mathrm{4NHM}}\big|_{\mathrm{diag}}
  =\frac{\big\langle\Phi_{1234}\big|I_{A}\big|\Phi_{1234}\big\rangle
        +\big\langle\Phi_{1234}\big|I_{B}\big|\Phi_{1234}\big\rangle}
        {M\,\big\langle\Phi_{1234}\big|\Phi_{1234}\big\rangle},
  \label{app:mr:collapsed}
\end{equation}
\noindent where we set $I_{A_{1234}} \equiv I_{A}$ and $I_{B_{1234}} \equiv I_{B}$ for notational convenience. 

In the reference state $\Phi_{1234}$, the inner moment $I_A$ depends only on
the internal coordinates of $A$ and the outer moment $I_B$ only on the external
ones. The expectation values therefore factorize,
\begin{align}
  \big\langle\Phi_{1234}\big|I_{A}\big|\Phi_{1234}\big\rangle
    &=\big\langle\Phi_A\big|I_A\big|\Phi_A\big\rangle\,
      \big\langle\Phi_B\big|\Phi_B\big\rangle ,\\
  \big\langle\Phi_{1234}\big|I_{B}\big|\Phi_{1234}\big\rangle
    &=\big\langle\Phi_B\big|I_B\big|\Phi_B\big\rangle\,
      \big\langle\Phi_A\big|\Phi_A\big\rangle ,
  \label{app:mr:factorize}
\end{align}

\noindent with $\langle\Phi_{1234}|\Phi_{1234}\rangle
   =\langle\Phi_A|\Phi_A\rangle\langle\Phi_B|\Phi_B\rangle$.

Substituting these into Eq.~(\ref{app:mr:collapsed}) and canceling the leftover
norm in each term gives
\begin{equation}
  R^2_{m,\mathrm{4NHM}}\big|_{\mathrm{diag}}
  =\frac{\big\langle\Phi_B\big|I_B\big|\Phi_B\big\rangle}
        {M\,\big\langle\Phi_B\big|\Phi_B\big\rangle}
  +\frac{\big\langle\Phi_A\big|I_A\big|\Phi_A\big\rangle}
        {M\,\big\langle\Phi_A\big|\Phi_A\big\rangle} .
  \label{app:mr:twoterms}
\end{equation}
Defining the matter radius of
the inner three-body subsystem and that of the effective two-neutron halo as
\begin{equation}
  R^2_{m,\mathrm{inner}}(k_A)=
  \frac{\big\langle\Phi_A\big|I_A\big|\Phi_A\big\rangle}
       {M_A\,\big\langle\Phi_A\big|\Phi_A\big\rangle}
\end{equation}

and

\begin{equation}
  R^2_{m,\mathrm{2NHM}}(k_B)=
  \frac{\big\langle\Phi_B\big|I_B\big|\Phi_B\big\rangle}
       {M\,\big\langle\Phi_B\big|\Phi_B\big\rangle},
  \label{app:mr:defs}
\end{equation}
respectively,
Eq.~(\ref{app:mr:twoterms}) gives
\begin{equation}
  R^2_{m,\mathrm{4NHM}}\big|_{\mathrm{diag}}
  =R^2_{m,\mathrm{2NHM}}(k_B)
   +\frac{M_A}{M}\,R^2_{m,\mathrm{inner}}(k_A) .
  \label{app:mr:exact}
\end{equation}

\subsection{Universal scaling and the limit}
\label{app:mr:limit}

In the unitary limit the three-body blocks carry no intrinsic length scale: the
universal wave function depends on $k$ only through the combination $kR$, so
dimensional analysis fixes every squared length to scale as $1/k^{2}$. Hence
\begin{align}
  R^2_{m,\mathrm{inner}}(k_A)&=\frac{\tilde c_A}{k_A^{2}}\ \,,
  \nonumber \\
  R^2_{m,\mathrm{2NHM}}(k_B)&=\frac{\tilde c_B}{k_B^{2}}\ \ ,
  \label{app:mr:twoblocks}
\end{align}
with $\tilde c_A,\tilde c_B$ pure numbers. Inserting Eq.~(\ref{app:mr:twoblocks})
into Eq.~(\ref{app:mr:exact}) gives
\begin{equation}
  \frac{R^2_{m,\mathrm{4NHM}}\big|_{\mathrm{diag}}}{R^2_{m,\mathrm{2NHM}}(k_B)}
  =1+\frac{M_A}{M}\,\frac{\tilde c_A}{\tilde c_B}\,\left(\frac{k_B}{k_A}\right)^{2}
  \;\xrightarrow[k_B/k_A \to0]{}\;1 .
  \label{app:mr:ratio}
\end{equation}
In the limit $k_B/k_A \to 0$ the inner cluster, of size $\sim 1/k_A$,
becomes a vanishing fraction of the outer halo, of size $\sim 1/k_B$. In
this precise sense the compact subsystem $A$ becomes point-like, and the
4NHM matter radius reduces to the 2NHM one.

\end{document}